\documentclass[10pt,preprint]{aastex}

\def\ltsima{$\; \buildrel < \over \sim \;$}
\def\gtsima{$\; \buildrel > \over \sim \;$}
\def\lsim{\lower.5ex\hbox{\ltsima}}
\def\gsim{\lower.5ex\hbox{\gtsima}}

\newdimen\minuswidth    %define @ width of minus sign for tables
\setbox0=\hbox{$-$}
\minuswidth=\wd0
\catcode`@=\active
\def@{\kern\minuswidth}
\setbox0=\hbox{\rm0}
\shorttitle{} 
\shortauthors{Ferraro et al.}
 
\begin{document} 
 \title{MIKiS: the Multi-Instrument Kinematic Survey of Galactic
   Globular Clusters. I. Velocity dispersion profiles and rotation
   signals of 11 globular clusters\footnote{Based on FLAMES and KMOS
     observations performed at the European Southern Observatory as
     part of the Large Programme 193.D-0232 (PI: Ferraro).}}

\author{
F. R. Ferraro\altaffilmark{2,3},
A. Mucciarelli\altaffilmark{2,3},
B. Lanzoni\altaffilmark{2,3}, C. Pallanca\altaffilmark{2,3}, E. Lapenna\altaffilmark{2,3},
L. Origlia\altaffilmark{3}, E. Dalessandro\altaffilmark{3},
E. Valenti\altaffilmark{4}, G. Beccari\altaffilmark{4}, M. Bellazzini\altaffilmark{3},
E. Vesperini\altaffilmark{5}, A. Varri\altaffilmark{6},
A. Sollima\altaffilmark{3}
}
\affil{\altaffilmark{2} Dipartimento di Fisica e Astronomia,
  Universit\`a degli Studi di Bologna, via Gobetti 93/2, I$-$40129
  Bologna, Italy}
\affil{\altaffilmark{3}INAF-Osservatorio di astrofisica e scienza
  dello spazio di Bologna, Via Gobetti 93/3, 40129, Bologna, Italy}
\affil{\altaffilmark{4}European Southern Observatory,
  Karl-Schwarzschild-Strasse 2, 85748 Garching bei M\"{u}nchen,
  Germany}
\affil{\altaffilmark{5} Dept. of Astronomy, Indiana University,
  Bloomington, IN, 47401, USA}
\affil{\altaffilmark{6} Institute for Astronomy, University of
  Edinburgh, Royal Observatory, Blackford Hill, Edinburgh EH9 3HJ, UK}
\date{23 April 2018}

\begin{abstract}
We present the first results of the Multi-Instrument Kinematic Survey
of Galactic Globular Clusters, a project aimed at exploring the
internal kinematics of a representative sample of Galactic globular
clusters from the radial velocity of individual stars, covering the
entire radial extension of each system. This is achieved by exploiting
the formidable combination of multi-object and integral field unit
spectroscopic facilities of the ESO Very Large Telescope. As a first
step, here we discuss the results obtained for 11 clusters from high
and medium resolution spectra acquired through a combination of FLAMES
and KMOS observations.  We provide the first kinematical
characterization of NGC 1261 and NGC 6496. In all the surveyed
systems, the velocity dispersion profile declines at increasing radii,
in agreement with the expectation from the King model that best fits
the density/luminosity profile.  In the majority of the surveyed
systems we find evidence of rotation within a few half-mass radii from
the center. These results are in general overall agreement with the
predictions of recent theoretical studies, suggesting that the
detected signals could be the relic of significant internal rotation
set at the epoch of the cluster's formation.
\end{abstract}
 

\section{INTRODUCTION}
\label{sec_intro}
Galactic Globular Clusters (GGCs) are the only astrophysical systems
that, within the time-scale of the age of the Universe, undergo nearly
all the physical processes known in stellar dynamics \citep{meylan+97,
  heggiehut03}.  Gravitational interactions among stars significantly
alter the overall energy budget and considerably affect the (otherwise
normal) stellar evolution, even generating exotic objects like blue
straggler stars, millisecond pulsars, X-ray binaries, and cataclysmic
variables (e.g., \citealt{bailyn95}).  Hence GGCs represent the ideal
laboratories where to study stellar dynamics and its effects on
stellar evolution.  Traditionally, GCs have been assumed to be
quasi-relaxed non-rotating systems, characterized by spherical
symmetry and orbital isotropy.  Hence, spherical, isotropic and
non-rotating models, with a truncated distribution function close to a
lowered-Maxwellian \citep[e.g.,][]{king66}, have been routinely used
to fit the observed surface brightness profiles and estimate the main
GC structural parameters, like the core and half-mass radii, the
concentration and even the total mass \citep{pryor+93, harris96,
  McLvdM05}.  While accurate cluster structural parameters start now
to be derived from a new generation of star density profiles (see
\citealt{ferraro+09,lanzoni+10, miocchi+13, dalessandro+13a}), this
information alone is not sufficient to univocally constrain the models
and get a comprehensive view of GC physics (\citealt{meylan+97}).  The
crucial missing ingredient is the information about cluster internal
dynamics.\footnote{Recent results suggest that insights on GC internal
  dynamics can be obtained from the observations of exotic stellar
  populations, like blue stragglers and millisecond pulsars
  \citep{ferraro+09, ferraro+12, lanzoni+16}.} In particular, a
detailed knowledge of the velocity dispersion (VD) profile and the
(possible) rotation curve of GGCs is still missing in the majority of
the cases.  This is essentially due to to observational difficulties.

In principle, VD and rotation can be obtained from different
approaches. One is to use the line broadening and the shift of
integrated-light spectra (e.g., \citealt{ibata+09, luz+11, luz+13,
  fabricius+14}). However, in the case of resolved stellar populations
like GGCs, this method can be prone to a severe ``shot noise bias''
\citep{dubath+97, lanzoni+13}. In fact, if a few bright giants bring a
dominant contribution into the integrated-light spectrum, the line
broadening provides a measure of their radial velocity (RV) scatter,
instead of a measure of the cluster VD due to the underlying stellar
population.  The other approaches consist in determining the cluster
VD from the velocities of statistically significant samples of
individual stars, either through resolved spectroscopy, thus obtaining
the line-of-sight VD (e.g., \citealt{lane+10, bellazzini+12,
  bianchini+13, husser+16, baum17, kamann+18, baum+18}), or via
internal proper motions (PMs), which provide the two VD components on
the plane of the sky (see, e.g., \citealp{bellini+14} for recent
results).  The latter is very challenging since it requires
high-precision photometry and astrometry on quite long time baselines
and it just started to be feasible, mainly thanks to the combination
of first and second epoch HST observations and the improved techniques
of data analysis \citep[see][and the GAIA Survey]{bellini+14,
  watkins+15, bellini+17}.  RVs are in principle easier to obtain
(through spectroscopy) and measurable in any cluster region and in GCs
at any distance from Earth within the Galaxy.  However, determining
the line-of-sight VD profile from individual stars over the entire
cluster extension is hard and very telescope time consuming, since it
requires to collect large samples of individual stellar spectra both
in environments of high stellar crowding (up to $\sim 7\times 10^5
L_\odot$ pc$^{-3}$; see \citealp{harris96}, 2010 version), and over
large sky regions (of $20\arcmin$-$40\arcmin$ diameter and even more).

To overcome these obstacles we recently proposed to combine
spectroscopic observations acquired from multiple instruments, with
different multi-object capabilities and different angular resolution
powers. In this paper we present first results obtained from the
proposed approach. Section \ref{sec_mikis} provides an overview of our
multi-instrument survey. In Section \ref{sec_obs} we describe the
observations and the adopted data reduction procedures. The
determination of the stellar RVs and the homogenization of the
different datasets is discussed in Section \ref{sec_RV}. The results
are presented in Section \ref{sec_resu}, while Sections
\ref{sec_discuss} and \ref{sec_conclu} are devoted to the discussion
and conclusions.

\section{The MIKiS survey}
\label{sec_mikis}
The Multi-Instrument Kinematic Survey of GGCs (hereafter the MIKiS
survey) was specifically designed to provide the entire VD and
rotation profiles of a representative sample of GGCs by fully
exploiting the spectroscopic capabilities available the ESO Very Large
Telescope (VLT).  The core scheme of the survey is to take advantage
of the specific characteristics of three different VLT spectrographs:
(1) the diffraction-limited integral field (IF) spectrograph SINFONI,
which allows to resolve GC stars in the innermost few arcseconds from
the center; (2) the seeing-limited IF spectrograph KMOS, which
provides an optimal coverage of the intermediate radial range (tens of
arcsecond scale), and (3) the wide-field multi-object spectrograph
FLAMES, which samples the external cluster regions (out to a dozen of
arcmin) with more than 100 fibers simultaneously.  This approach was
first tested, as a proof of concept, for the case of NGC 6388
\citep{lanzoni+13, lapenna+15a}.

\begin{figure}
\centering
\includegraphics[width=15cm]{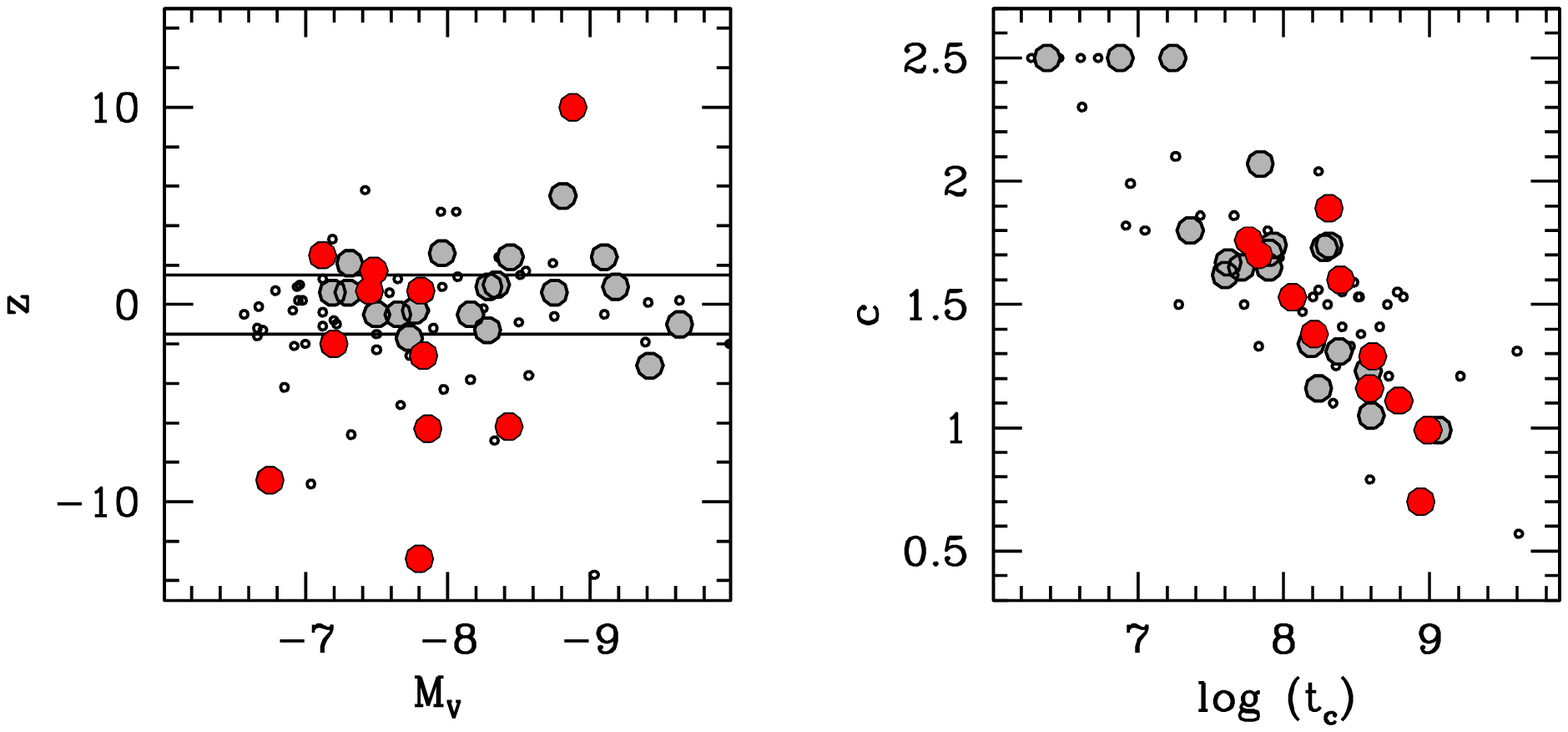}
\caption{Distribution of the GGCs observed in the MIKiS survey (large
  circles) in the planes of height on the Galactic plane vs. absolute
  integrated V-band magnitude {\it (left)} and concentration parameter
  vs. core relaxation time {\it (right)}. The 11 clusters discussed
  here are in red. The entire GGC population is also plotted for
  reference (small dots).}
\label{fig_sample}
\end{figure}

For the MIKiS survey we selected a sample of 30 GCs well
representative of the overall Galactic population (see Figure
\ref{fig_sample}): they properly encompass $(i)$ the cluster
dynamically-sensitive parameter space (spanning a large range of
central densities and a factor of 3 in the concentration parameter),
$(ii)$ different stages of dynamical evolution (the sample includes
both pre- and post- core-collapse GCs, with the core relaxation time
spanning almost 3 orders of magnitude), and $(iii)$ different
environmental conditions (they are distributed at different heights on
the Galactic plane, thus sampling both the bulge/disk and the halo
populations: $|z|<1.5$ kpc, and $1.5<|z|<13$, respectively). The
selected targets are also more luminous than $M_V=-6.8$ (i.e. populous
enough to guarantee large samples of giant stars for a meaningful
determination of the VD), relatively close to Earth (within $\sim 16$
kpc, thus providing spectra with good signal-to-noise ratios for stars
down to the sub-giant branch, in reasonable exposure times) and not
extremely metal-poor ([Fe/H]$> -1.8$, thus allowing RV measurements
with an accuracy of a few km/s also from relatively low-resolution IR
spectra).

Thanks to the adopted strategy and the selected cluster sample, the
MIKiS survey is expected to provide the full characterization of the
line-of-sight internal kinematics from the innermost to the outermost
regions of each clusters, with crucial impact on many hot-topics of GC
science.  We will properly search for signatures of systemic rotation
and intermediate mass ($10^3-10^4 M_\odot$) black holes, thus
providing new crucial insights on the physics and formation processes
of both GCs and these elusive compact objects \citep[see,
  e.g.,][]{baum05, miocchi07, varri+12, zocchi+17, tiongco+17}.  We
will also accurately determine the whole mass distribution and the
global amount of dark remnants (white dwarfs, neutron stars, stellar
mass black holes) in the sampled clusters.  While a complete census of
these stars is beyond any observational possibility, the VD profile is
sensitive to the whole mass enclosed in a stellar orbit. Hence, the
simultaneous knowledge of the density and the VD profiles can provide
reliable estimates of the stellar densities, mass-to-light ratios, and
cluster total mass \citep{mandushev+91, pryor+93, meylan+97, lane+10,
  sollima+12, zocchi+12, baum17, baumsoll17}.  We also aim at
characterizing the kinematics of multiple populations with different
light-element content, to provide crucial constraints to GC formation
scenarios (e.g., \citealt{vesperini+13, richer+13, bellini+15,
  henault+15, cordero+17}).  Finally, the MIKiS survey will allow the
exploration of GC dynamics in the proximity of their tidal limitation,
which is essential in order to formulate more realistic descriptions
of this class of stellar systems \citep[e.g.,][]{davoust77, McLvdM05,
  gieles+15}, to pin down the physical origin of recently claimed
``extra-tidal'' structures (e.g., see \citealp{olszewski+09,
  correnti+11} and, more recently, \citealp{kuzma+16, kuzma+18,
  carballo-bello+18}), to investigate the interplay with the external
tidal field \citep[e.g.,][]{heggie+95, varri+09}, and even to assess
the implications of more exotic possibilities such as small dark
matter haloes \citep[e.g.,][]{mashchenko+05, shin+13, penarrubia+17},
or modifications of the theory of gravity \citep[e.g.,][]{ibata+11,
  hernandez+13}.

While a few works based on the data acquired with the MIKiS survey
have been already published on specific sub-topics \citep{lapenna+15b,
  ferraro+16, sollima+16}, this is the first paper of a series
specifically devoted to discuss the kinematic results of the
survey. As a first step, here we present the VD profiles and the
rotation signals detected in the intermediate/outer regions of 11 GGCs
(highlighted in red in Figure \ref{fig_sample}), which represent the
bulk of the targets observed only with the FLAMES+KMOS combination.
In a series of future studies we will discuss the most intriguing
cases obtained from the combined use of the all three instruments
(SINFONI+KMOS+FLAMES).
 
\section{Observations and data reduction}
\label{sec_obs}
Within the MIKis survey we used FLAMES (\citealt{pasquini+00}) in the
GIRAFFE/MEDUSA combined mode (consisting of 132 deployable fibers
which can be allocated within a $25\arcmin$-diameter field of view),
adopting the HR21 grating setup, with a resolving power R$\sim 16200$
and a spectral coverage from 8484 \AA{} to 9001 \AA. This grating
samples the prominent Ca II triplet lines, which are excellent
features to measure RVs also in relatively low ($\sim 10$-15)
signal-to-noise ratio (SNR) spectra.  The target stars have been
selected from optical wide-field photometric catalogs presented in
previous papers (see e.g. \citealt{ferraro+04, lanzoni+07a,
  lanzoni+07b, dalessandro+13a, dalessandro+13b, ferraro+12}). We
selected only red giant branch stars brighter than $I=18.5$ and, in
order to avoid spurious contamination from other sources within the
fibers, we also requested that no bright neighbors (I$_{{\rm
    neighbor}}<{\rm I_{{\rm star}}}+1.0$) were present within
$2\arcsec$ from each target.  On average, 3-4 pointings have been
performed in each cluster.  For each pointing, multiple exposures,
with total integration times ranging from 900 s to 3000 s according to
the magnitude of the targets, were secured (see Table
\ref{tab_obs}). This provided SNRs$\sim$30 at the faintest magnitudes.
For each target cluster, one pointing has been dedicated to re-observe
10-20 stars in common with the pre-existing datasets that we retrieved
from the ESO archive (see Table \ref{tab_datasets}) in order to
increase at most the sample of individual FLAMES spectra.  The data
reduction was performed by using the FLAMES-GIRAFFE
pipeline\footnote{http://www.eso.org/sci/software/pipelines/},
including bias-subtraction, flat-field correction, wavelength
calibration with a standard Th-Ar lamp, re-sampling at a constant
pixel-size and extraction of one-dimensional spectra.  Typically 15-20
fibers were used to measure the sky in each exposure. These spectra
have been averaged to obtain a master sky spectrum, which was then
subtracted from each target spectrum.

We have used KMOS \citep{sharples+10} to measure red giant stars with
$J<14$, located within $\sim 70\arcsec$ from each cluster center.
KMOS is a spectrograph equipped with 24 deployable IF units that can
be allocated within a $7.2\arcmin$ diameter field of view.  Each IF
unit covers a projected area on the sky of about
$2.8\arcsec\times2.8\arcsec$, sampled by an array of 14$\times$14
spatial pixels (spaxels) with an angular size of $0.2\arcsec$ each. We
have used the YJ grating covering the 1.00-1.35 $\mu$m spectral range
at a resolution R$\approx$3400, corresponding to a sampling of about
1.75 $\rm \mathring{A}$ pixel$^{-1}$, i.e. $\sim$ 46 km s$^{-1}$
pixel$^{-1}$ at 1.15 $\mu$m.  This instrumental setup is especially
effective in simultaneously measuring a number of reference telluric
lines in the spectra of giant stars, for an accurate calibration of
the RV, despite the relatively low spectral resolution. Typically, 7-8
pointings have been secured in each cluster.  The total on-source
integration time for each pointing was 3-5 min and it has been
obtained with three sub-exposures of 60-100 s each, dithered by
$0.2\arcsec$ for optimal flat-field correction. The typical SNR of the
observed spectra is $\gtrsim$ 50.  We used the ``nod to sky'' KMOS
observing mode and nodded the telescope to an off-set sky field at
$\approx 6\arcmin$ North of the cluster center, for a proper
background subtraction. The spectroscopic targets have been selected
from near-IR catalogs published by our group
\citep{ferraro+00,valenti+04,valenti+07}, on the basis of their
position in the color-magnitude diagrams.  We selected rad giant
targets with $J < 14$ mag and with no stars brighter than $J=15$
within $1\arcsec$ from their center.  We also used ACS-HST data in the
F606W and F814W bands, from \citet{sarajedini+07}, to identify
additional stars not present in the IR catalog. The raw data have been
reduced using the KMOS pipeline$^3$ which performs background
subtraction, flat field correction and wavelength calibration of the
2D spectra.  The 1D spectra have been extracted manually by visually
inspecting each IF unit and selecting the spectrum from to the
brightest spaxel in correspondence of each target star centroid, in
order to minimize the effects of possible residual contamination from
nearby stars and/or from the unresolved stellar background. Normally,
one star was measured in each IF unit. Only in a few cases two or more
resolved stars were clearly distinguishable in a single KMOS IF unit,
and their spectra were extracted.

\section{Radial velocity measurements}
\label{sec_RV}
To measure the RVs of the target stars we cross-correlated the
observed spectra (corrected for heliocentric velocity) with a template
of known velocity, following the procedure described in
\citet{tonry+79} and implemented in the FXCOR software under IRAF.  As
templates we used synthetic spectra computed with the SYNTHE code
\citep[see e.g.][]{sbordone+04}, adopting the clusters metallicity and
appropriate atmospheric parameters according to the evolutionary stage
of the targets. All the synthetic spectra have been convoluted with a
Gaussian profile to reproduce the spectral resolution of each dataset.
Finally, a visual inspection of all the observed spectra shifted to
zero velocity, compared with the synthetic template, has been
performed to assess the quality of the solution.

For the FLAMES targets we measured the RV in three different regions
of the same spectrum (region 1: 8490\AA$<\lambda<8630$\AA, region 2:
8630\AA$<\lambda<8770$\AA, region 3: 8790\AA$<\lambda<8900$\AA), each
including a large number of lines.  The star velocity and its
uncertainty are then obtained, respectively, as the mean of these
different measures, and their dispersion divided by the square root of
the number of spectral regions used.  The typical uncertainties in the
RVs derived for FLAMES targets are of the order of 0.1-0.3 km
s$^{-1}$.  Before combining RV measures obtained from different FLAMES
gratings, we checked for possible systematic offsets. Since we adopted
the RVs acquired in the MIKiS survey as reference, for the stars
observed with the HR21 setup we checked the accuracy of the zero-point
of the wavelength calibration by measuring the position of several
emission sky lines available in the spectral range, finding no
significant offsets.  Then, in order to align the other FLAMES
datasets we used the stars in common (typically a dozen for each
cluster), always finding a very good agreement.  When multiple
exposures were available for the same star, we first verified that RV
measures agreed within the errors (if not, the star was assumed to be
a candidate binary system and excluded from the
analysis),\footnote{Operationally, we determined the scatter of the RV
  measures available for a given target and compared it to the rms of
  the error distribution of the stars with similar magnitude: if it
  was larger by a factor of 3 or more, then the target was classified
  as candidate binary and excluded from the analysis. Clearly, this is
  just a zero-order selection and some binary system could still
  present in our samples. However, our observations are limited to the
  brightest portion of the color-magnitude diagram, essentially
  sampling the red giant branch, and we therefore expect no
  significant effects on the discussed results: in fact, binaries with
  red giant companions in GCs amount to a very small fraction ($\lsim
  2\%$) and they show RV variations of just 1-2 km s$^{-1}$ (see,
  e.g., \citealp{sommariva+09}).}  and we then determined its final RV
as the weighted mean of all the measures, by using the individual
errors as weights.

For the KMOS targets the precision on the derived RVs has been
estimated through Montecarlo simulations, using cross-correlation
against 500 synthetic spectra of a given SNR per pixel.  The synthetic
spectra have been calculated over the wavelength region covered by
KMOS and assuming the appropriate metallicity of the cluster and the
typical atmospheric parameters of the observed targets.  Each
synthetic spectrum has been resampled at the KMOS pixel-scale (1.75
pixel/\AA) and a Poissonian noise has been injected to reproduce a
given SNR per pixel (we simulated SNRs between 20 and 80).  The
dispersion of the derived RV distribution is adopted as the 1-$\sigma$
uncertainty for a given SNR.  An exponential relation between the SNR
and the RV precision as estimated from the above procedure has been
derived and used to attribute a RV error to each target.  The final
errors for KMOS RVs typically are of the order of 1-5 km s$^{-1}$.

To homogenize the RV measures obtained from FLAMES and KMOS, we used
at least 10 targets per cluster that have been observed with both the
spectrographs. The measured offsets (3-5 km s$^{-1}$) can be explained
by a combined effect of the relatively low spectral resolution of KMOS
and the low metallicity of a few clusters (yielding to KMOS spectra
with lower SNR). We then double-checked the realignment of the two
datasets by comparing the systemic velocities (see Section
\ref{sec_vsys}) of the FLAMES and the KMOS samples separately in each
GC. Finally, for each star in common between the two samples, we
adopted the RV measure obtained from the (higher resolution) FLAMES
spectra.

\section{Results}
\label{sec_resu}
\subsection{Systemic velocities}
\label{sec_vsys}
\begin{figure}
\includegraphics[width=15cm]{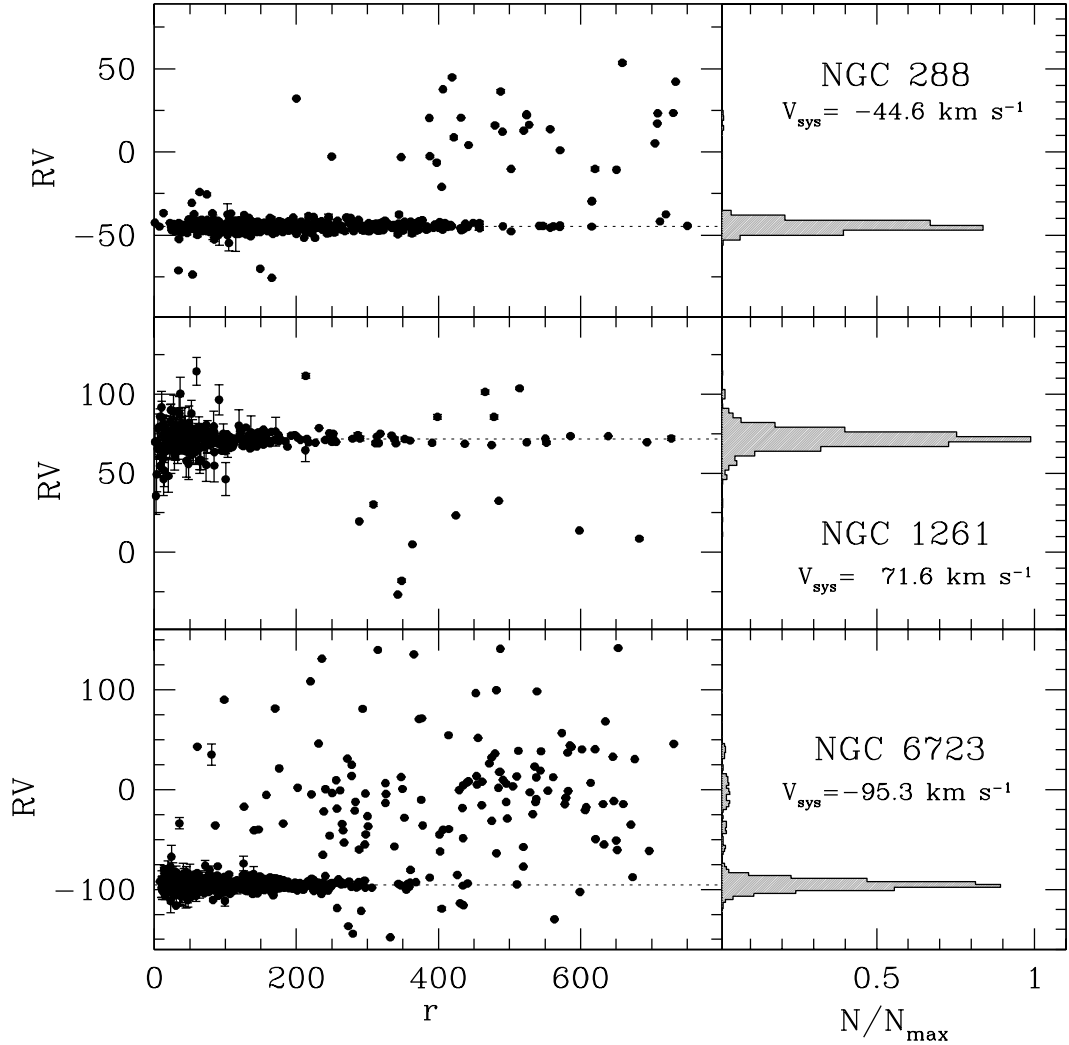}
\caption{{\it Left panels:} Radial velocities as a function of the
  distance from the cluster center obtained from the KMOS+FLAMES
  observations of NGC 288, NGC 1261 and NGC 6723 (see labels). RVs are
  in km s$^{-1}$, radial distances are in arcseconds. {\it Right
    panels:} histogram of the corresponding RV distribution, with
  value of the derived systemic velocity labelled for each
  cluster. The histograms are normalized to their peak values.}
\label{fig_vrr1}
\end{figure}

\begin{figure}
\includegraphics[width=15cm]{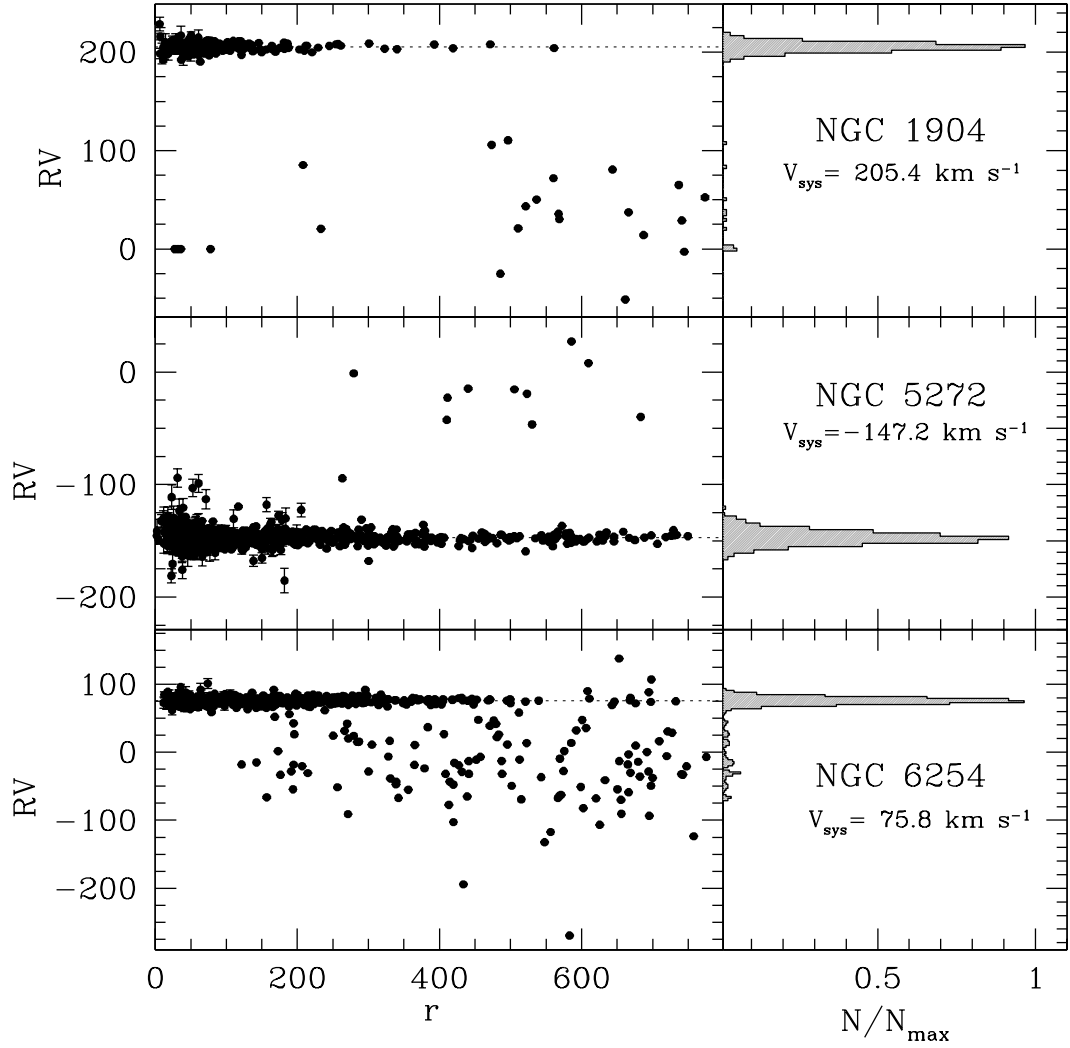}
\caption{As in Figure 1, but for NGC 1904, NGC 5272 and NGC 6254.}
\label{fig_vrr2}
\end{figure}

\begin{figure}
\includegraphics[width=15cm]{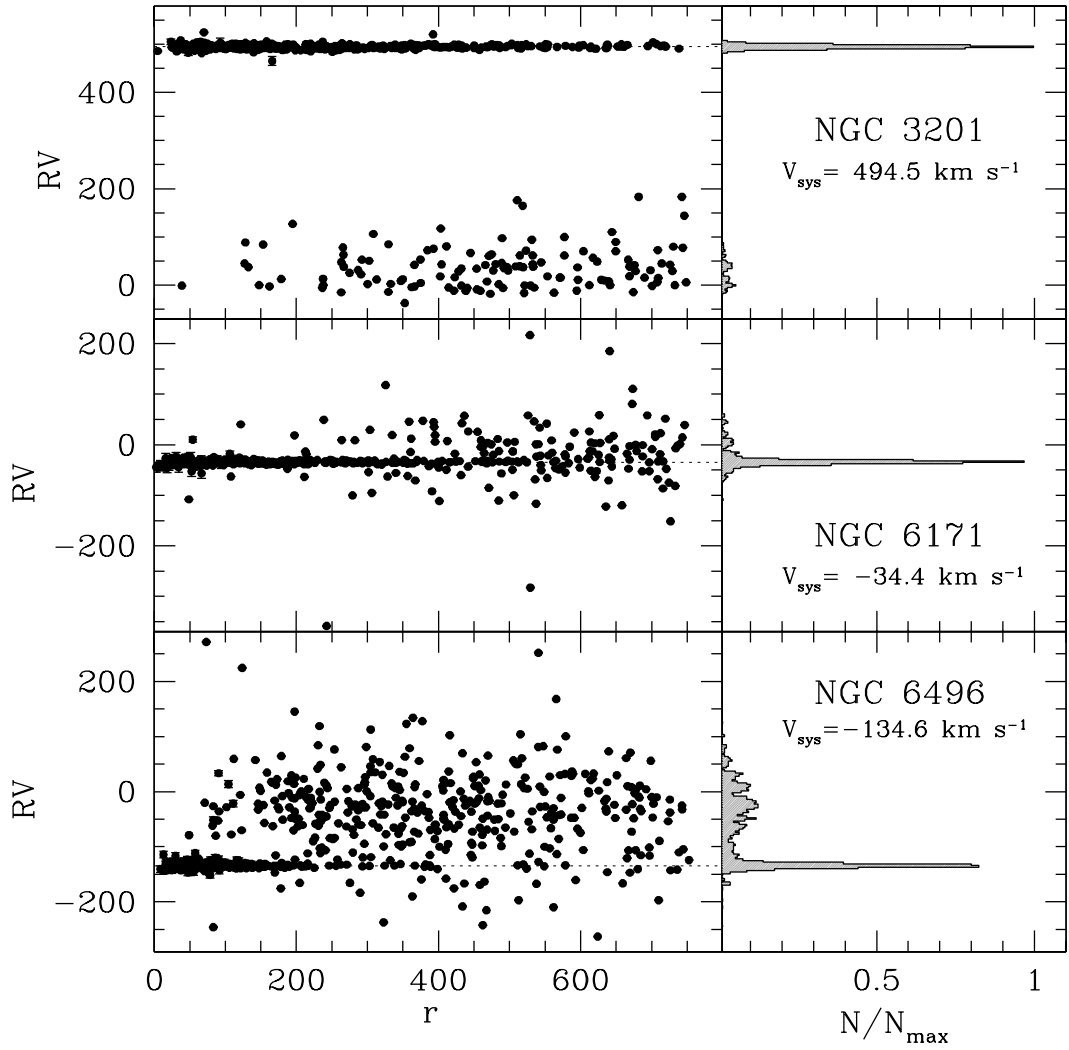}
\caption{As in Figure 1, but for NGC 3201, NGC 6171 and NGC 6496.}
\label{fig_vrr3}
\end{figure}

\begin{figure}
\includegraphics[width=15cm]{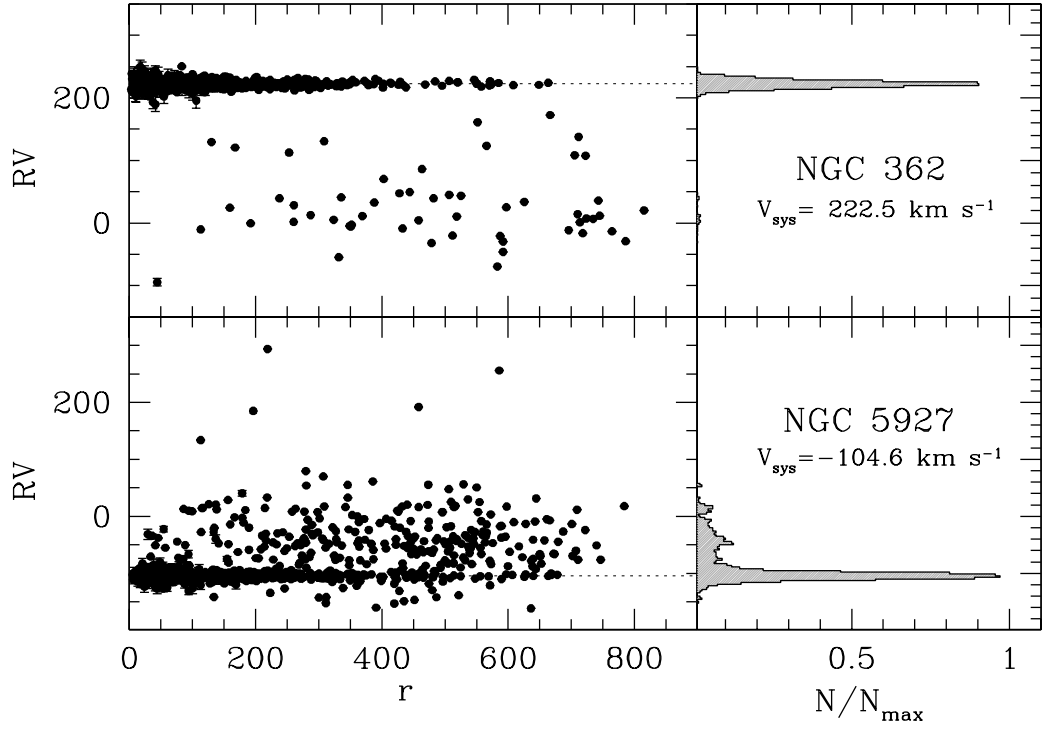}
\caption{As in Figure 1, but for NGC 362 and NGC 5927.}
\label{fig_vrr4}
\end{figure}

The total number of stars with measured RV in each program cluster is
listed in Table \ref{tab_nstar}, together with the minimum and the
maximum distance from the center sampled by the collected datasets.
The catalogs of the measured RVs are freely downloadable at the
  MIKiS web page\footnote{
    http://www.cosmic-lab.eu/Cosmic-Lab/MIKiS\_Survey.html}.
Figures \ref{fig_vrr1}--\ref{fig_vrr4} show the distribution of the
measured RVs as a function of the distance from the cluster center.
The data span a large range of radial distances from the central
regions out to $\sim 800\arcsec$, extending, in some cases, even
beyond the nominal cluster tidal radius. The population of cluster
members is clearly distinguishable as a narrow, strongly peaked
component, which dominates the sample at radii smaller than $\sim
500\arcsec$.

In a few clusters, the field component is clearly identified as a
broad distribution at all sampled radii, homogeneously spanning a wide
range of RVs (typically $\sim 200$ km s$^{-1}$). Three systems (namely
NGC 5927, NGC 6171 and NGC 6496) appear to be particularly affected by
the tail of the field velocity distribution, which significantly
overlaps that of the cluster.  While this can impact the determination
of the cluster VD and rotation (see Sections \ref{sec_vdisp} and
\ref{sec_vrot}), it is not an issue for measuring the systemic
velocity of the cluster ($V_{\rm sys}$). In fact, this latter has been
obtained by conservatively considering only stars with RVs in a
relatively narrow strip of values (typically $\Delta RV=\pm 20$km
s$^{-1}$) centered at the histogram peak velocity. We also excluded
the most external radial bins, where the number of field stars can be
comparable to (or even larger than) that of cluster members.  Assuming
that the RV distribution is Gaussian, we used a Maximum-Likelihood
method \citep[e.g.,][see also Section \ref{sec_vdisp}]{walker06} to
estimate its mean and uncertainty.  The values of $V_{\rm sys}$ thus
obtained for each cluster are listed in Table \ref{tab_vsys_sig0} and
labelled in Figures \ref{fig_vrr1}--\ref{fig_vrr4}.  In the following,
we wil use $\widetilde V_r$ to indicate RVs referred to the cluster
systemic velocity: $\widetilde V_r\equiv V_r-V_{\rm sys}$.

\subsection{Cluster membership}
\label{sec_memb}
As discussed above, the cluster membership selection is
straightforward in all the sampled systems but three (namely NGC 5927,
NGC 6171 and NGC 6496).  Thus, before discussing the procedure adopted
to determine the VD profile and to search for signals of systemic
rotation, we describe the approach used to decontaminate these three
systems.  In principle, beside the RV values, an additional constrain
to the cluster membership can be obtained from the stellar
metallicity, provided that the two samples (cluster and Galactic
field) have different [Fe/H] distributions.  KMOS spectra can provide
metallicities with uncertainties of about 0.2-0.3 dex, due to the low
spectral resolution of the instrument (and the low metallicities of
most of the program cluster). Moreover, the KMOS targets are located
in the innermost cluster regions, where the field contamination is not
critical.  Hence, metallicities from KMOS spectra have not been used
to distinguish cluster from field stars.  Instead, for all the targets
observed with the HR21 grating of FLAMES in the three most
contaminated systems, we have estimated the [Fe/H] ratio from the Ca
II triplet lines.  The equivalent width has been measured by fitting
the Ca II triplet profile with a Voigt function. Then, [Fe/H] values
have been derived for most of the targets by adopting the relation of
\citet{vasquez15}, which is calibrated as a function of the $K$-band
magnitude.  For the stars with no $K$ magnitude information we used
the relation of \citet{carrera+07}, which is calibrated in the
$V$-band.  The reference horizontal branch magnitudes adopted in this
analysis are from \citet{harris96}.  We checked that the two
considered relations are consistent and the derived metallicities are
compatible within the measure uncertainties ($\sim$0.1-0.12 dex).

\begin{figure}
\includegraphics[width=15cm]{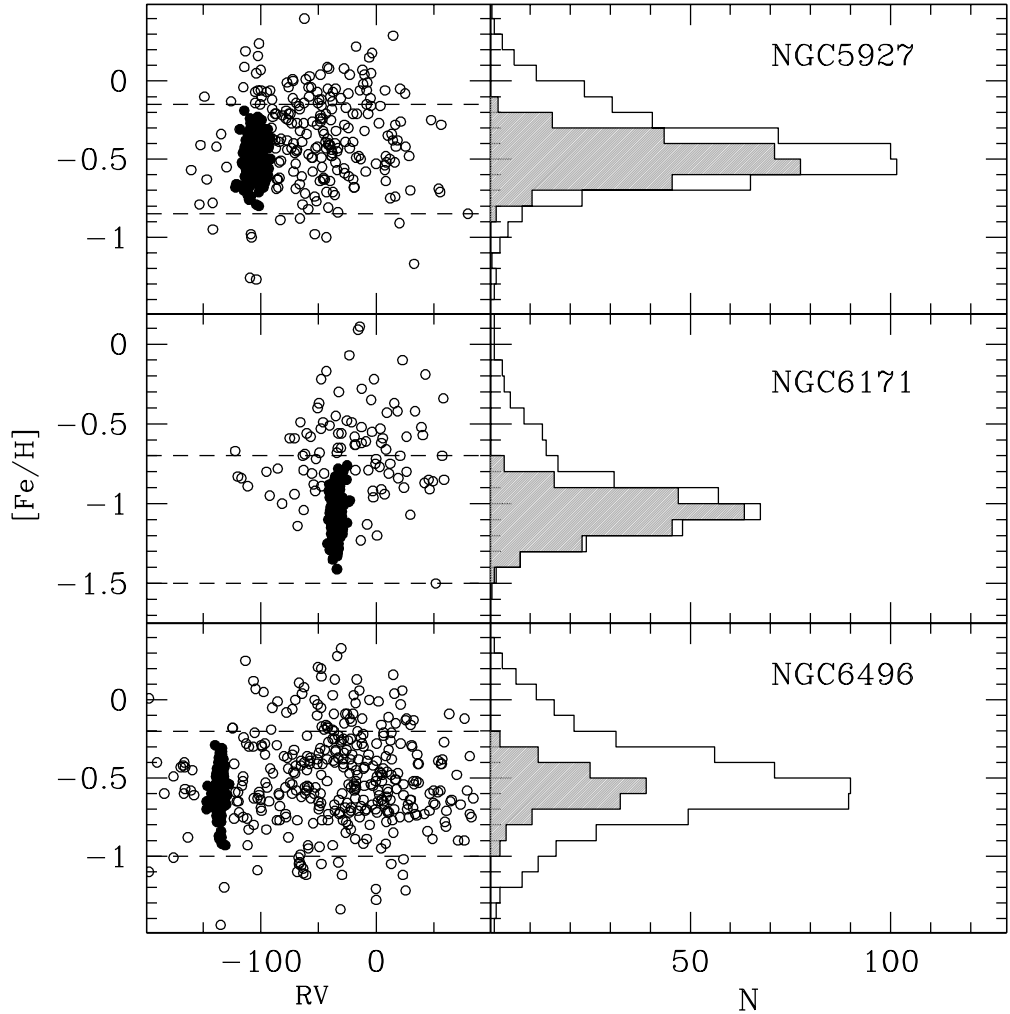}
\caption{[Fe/H] abundance ratios as a function of radial velocities
  ({\it left panels}) and corresponding metallicity distributions
  ({\it right panels}) for the targets observed with the HR21 grating
  of FLAMES in the three clusters with pronounced Galactic field
  contamination.  In each panel, the range of cluster metallicities is
  delimited by the two dashed lines.  The targets with the highest
  probability to be cluster members are plotted as filled circles and
  their corresponding metallicity histogram is shaded in grey.}
\label{fig_vrfe}
\end{figure}

Figure \ref{fig_vrfe} shows the measured metallicity as a function of
RV.  As apparent, cluster members clump in restricted ranges of RV and
[Fe/H] values (in agreement with the literature, the metallicities are
centered at [Fe/H]=$-0.49$, $-1.06$, and $-0.56$ in NGC 5927, NGC 6171
and NGC 6496, respectively, and have intrinsic dispersions of the same
order of the uncertainties; e.g., \citealp{harris96}).  Instead, field
stars have much larger scatters.  Unfortunately, the metallicity
distribution of the field largely overlaps that of the three clusters
(the most favorable case being NGC 6171), and it therefore does not
allow to implement a conclusive separation between the two
components. We therefore performed just a first-order decontamination
by excluding from the following kinematical analysis all the stars
with metallicity below and above the cluster values (which are marked
by the two dashed lines in each panel of Figure \ref{fig_vrfe}).
Clearly, this leaves a sample of stars (either having [Fe/H]
compatible with the cluster value, or with no metallicity information)
that still include field contaminants, with a RV distribution
partially overlapping that of genuine cluster members.  To remove such
a residual field contamination, we thus adopted the double-Gaussian
statistical approach described below.
 
\subsection{Velocity dispersion profile}
\label{sec_vdisp}
The projected VD profile\footnote{Formally, the derived values of
  $\sigma_P(r)$ are the second moments of the RV distribution, which
  coincide with the true VD only in the case of null
  rotation. However, as discussed below (Section \ref{sec_discuss}),
  the rotational velocity always provides a negligible contribution,
  and this measure can therefore be assumed as the true stellar VD.},
$\sigma_P(r)$, has been determined from the measured RVs by splitting
the surveyed area in a set of concentric annuli, chosen as a
compromise between a good radial sampling and a statistically
significant number ($\gsim 40$) of stars.\footnote{The number of stars
  per bin can be lower than 40 in the most external annulus of a few
  clusters because of the intrinsic outward density decline, and in
  NGC 1904 because only 235 RVs in total are available for this
  system.}  In each radial bin, obvious outliers (i.e., stars with RVs
in clear disagreement with the cluster distribution in that radial
interval) have been excluded from the analysis and a
$3\sigma$-clipping selection about the cluster systemic velocity has
been used to further clean the sample. Then, $\sigma_P(r)$ has been
computed from the dispersion of the remaining $\widetilde V_r$ values
following the Maximum-Likelihood method described in \citet[][see also
  Martin et al. 2007; Sollima et al. 2009]{walker06}.  Errors have
been estimated following \citet{pryor+93}.

For the three most contaminated clusters we assumed that the RV
distribution of the stars survived to the metallicity selection (see
Section \ref{sec_memb}) is the combination of two Gaussian functions,
one representing the cluster contribution, the other corresponding to
the field. Obviously, the cluster Gaussian is peaked at $V_{\rm sys}$
and its dispersion varies from one bin to another following the VD
profile.  Instead, the Gaussian function corresponding to
foreground/background stars in the direction of each cluster is peaked
at the characteristic velocity of the Galactic field in that region,
and has a much larger dispersion. To determine the peak and the
dispersion values of the field Gaussian we used the RV distribution of
the stars observed at large distances from the cluster centre, where
the field is largely dominant with respect to the cluster.
Under the assumption that the observed RV distribution has a double
Gaussian profile, we thus used the same Maximum-Likelihood method
described above to determine the cluster VD from the dispersion of the
cluster Gaussian function \citep[see][for more details]{sollima+16}.

The resulting VD profiles are shown in Figure \ref{vdisp} and listed
in Table \ref{tab_vdisp}. For NGC 1261 and NGC 6496 these are the
first determinations in the literature.  Three clusters (namely NGC
362, NGC 3201, NGC 6254) are in common with the sample recently
published by \citet{kamann+18}, who used the IF spectrograph MUSE to
survey the central regions of 25 GCs.  The comparison between the VD
profiles obtained in the present work and in \citet{kamann+18} for
these three systems (Figure \ref{fig_confro_kam}) shows a good
agreement in the radial region in common, and clearly illustrates that
the two studies are well complementary, with the MUSE data covering
distances out to $\sim 50\arcsec$, while our KMOS+FLAMES spectra
extending the VD profiles out to more than $500\arcsec$.  This nicely
demonstrates that a proper multi-instrument approach is able to
provide the VD profile of Galactic GCs over the entire radial
extension of each system.
   
\subsection{Systemic rotation signals}
\label{sec_vrot}
The spatial distribution of the surveyed stars in the plane of the sky
is symmetric with respect to the cluster center out to a maximum
distance ($d_{\rm max}$) that varies from a system to another, but is
always larger than twice the half-mass radius ($r_h$). This allowed us
to search for evidence of systemic rotation over a significant portion
of the radial extension of each cluster.

For this purpose, we used the method fully described in
\citet{bellazzini+12}. In short, we considered a line passing through
the cluster center with position angle (PA) varying from $0\arcdeg$
(North direction) to $180\arcdeg$ (South direction), by steps of
$10\arcdeg$ and with $90\arcdeg$ direction corresponding to the
East. For each value of PA, such a line splits the observed sample in
two.  If the cluster is rotating along the line-of-sight, we expect to
find a value of PA that maximizes the difference between the median
RVs of the two sub-samples ($\Delta\widetilde V_{\rm med}$), since one
component is mostly approaching, while the other is receding with
respect to the observer.  Moving PA from this value has the effect of
gradually decreasing the difference in median RV.  Hence, the
appearance of a coherent sinusoidal behavior of $\Delta\widetilde
V_{\rm med}$ as a function of PA is a signature of rotation and its
best-fit sine function provides an estimate of the rotation amplitude
($A_{\rm rot}$) and the position angle of the cluster rotation axis
(PA$_0$).  In the presence of systemic rotation, the stellar
distribution in a diagram showing the measured RVs ($\widetilde V_r$)
as a function of the projected distances from the rotation axis (XR)
shows an asymmetry, with two diagonally opposite quadrants being more
populated than the remaining two.  In combination with these diagrams,
we also used three different estimators to quantify the statistical
significance of any detected signal.  We performed a
Kolmogorov-Smirnov test to quantify the probability that the RV
distributions of the two sub-samples on each side of the rotation axis
(i.e., one having positive values of the rotated coordinate XR, the
other one having XR$<0$) are extracted from the same parent
distribution.  We then used both the Student's t-test and a
Maximum-Likelihood approach for assessing the statistical significance
of the difference between the two sample means. The first method has
the advantage of being non-parametric, while the other two assume that
the data have normal distributions (which is reasonable for samples of
stellar RVs in a GC).

To investigate the presence of ordered motions, we considered only the
stars used to determine the VD profile (i.e., we neglected all the
outliers and field contaminants excluded from the previous analysis,
and for the three most contaminated GCs we limited the search to
distances with negligible field contamination).  We searched for
rotation signals over each system as a whole, and also in discrete
radial bins out to the maximum distance allowed $d_{\rm max}$. In the
case of weak rotation, this should allow to detect at least the
maximum of the signal (i.e., the rotational velocity peak), which is
expected to be at some non-vanishing distance from the rotation axis.
While no significant evidence of global rotation has been found when
the stars observed over the entire cluster extension are considered,
in all cases we were able to identify the radial region with the
strongest rotation signal.  The results are summarized in Figures
\ref{fig_vrot1}-\ref{fig_vrot4} and in Table
\ref{tab_vrot}.\footnote{Note that the probabilities and
    significance levels listed in Table \ref{tab_vrot} could be
    slightly overestimated because the statistical rejection of the
    null hypothesis (no rotation) is more likely when multiple
    hypotheses are tested (in our approach we searched for rotation
    signatures in a few radial intervals per cluster). A way to take
    this into account \citep{bonferroni36} is to multiply the KS
    probabilities listed in Table \ref{tab_vrot} by the number of bins
    surveyed (see the last column of the table). However, given the
    small number of bins used and the high-significance level of the
    detected signals, this correction does not significantly alter our
    results. On the other hand, the crude application of the
    Bonferroni correction to this scientific case is not completely
    appropriate, since the probability of detecting rotation is not
    the same in all radial bins.  In fact, the amplitude of the
    rotation curve is expected to have a maximum (hence to be more
    easily detectable) at some off-centered radius (see, e.g., Fig. 10
    in \citealp{kacharov+14} and Fig. 3 in \citealp{tiongco+17}), and
    the application of the Bonferroni approach possibly over-corrects
    (artificially decreases) the significance of the detections in
    this case.}  Indications of systemic rotation have been found in
the majority (10 out of 11) of the GCs in our sample. The most
significant (at more than 3-$\sigma$) signals are detected in six
cases (NGC 288, NGC 362, NGC 1904, NGC 3201, NGC 5272 and NGC 6171),
while we estimate a $\sim 2$-$\sigma$ statistical significance for
NGC 1261 (although the number of stars is relatively small), NGC
5927, NGC 6496 and NGC 6723, and no evidence of systemic rotation in
NGC 6254.

\section{Discussion}
\label{sec_discuss}
We presented the first results obtained from the MIKiS survey from the
analysis of $\sim 6275$ high/medium resolution spectra of individual
stars sampling the entire radial extension of 11 GGCs.  This dataset
allowed us to accurately determine the systemic velocity and VD
profile, and to investigate the presence of ordered rotation in each
system. In particular, we provided the first determination of the
internal kinematical properties of NGC 1261 and NGC 6496.

%-----> Vsys
For the majority of the clusters the derived systemic velocities are
in very good agreement with the results published in the literature
(see Table \ref{tab_vsys_sig0} for the comparison with the values
quoted in the \citealp{harris96} catalog; see also \citealt{lane+10},
\citealt{kimmig+15}, \citealt{lardo+15}).  The only notable exception
is NGC 6496, for which we find a difference of $\sim 20$ km s$^{-1}$
with respect to the value quoted in the Harris catalog. This can be
explained by noticing that the latter is determined from very few
(less than 10) RV measures only (the most recent source being
\citealp{rutledge+97}, who acquired four spectra in this cluster),
while we used more than 100 stars. We can thus confidently conclude
that the systemic velocity quoted here for NGC 6496 is the most
accurate and reliable so far.

%-----> rotation
The MIKis survey collections of RVs provide a symmetric sampling of
the plane of the sky around each cluster center, out to $\sim 2$-7
$r_h$ depending on the system.  We have thus been able to search for
signatures of systemic rotation over a significant radial portion of
each GC, finding that the majority of the targets (9 over 11) show
evidence of rotation at intermediate cluster-centric distances (see
Table \ref{tab_vrot}).  This is in agreement with the findings of
\citet{kamann+18}, who, from the analysis of MUSE spectra, concluded
that $\sim 60\%$ of the GCs in their sample presents by some degree of
internal rotation. The dataset presented here offers the advantage of
a significantly larger spatial coverage. In fact, for the three
clusters in common (NGC 362, NGC 3201, and NGC 6254), the MUSE
observations cover 1, 0.5 and 0.8 $r_h$, respectively, while our data
extend much further out (to $\sim 5.4, 3.2$ and 3.2 $r_h$,
respectively). \citet{kamann+18} found no signatures of rotation in
the latter two GCs, while a marginal signal was detected in the centre
(at $\sim 0.05 r_h$) of NGC 362.  We confirm the absence of rotation
in NGC 6254, and we find strong signatures both in NGC 3201 (at $\sim
2 r_h$) and in NGC 362 (at $4.2 r_h$).  These results, however, are
not in disagreement, since our detections are well outside the regions
sampled by the MUSE observations of \citet{kamann+18}. Detailed
comparisons with other previous works in the literature are not
straightforward because global rotation amplitudes, determined all
over the radial range sampled by the observations, are usually
quoted. However, we can notice that, in agreement with our results,
some signatures of systemic rotation were already detected in NGC 288,
NGC 362, NGC 1904, NGC 5272, and NGC 6171 \citep[see][and references
  therein]{lane+10, scarpa+11, bellazzini+12, kimmig+15, lardo+15}.

The observations collected so far (this paper, and, e.g.,
\citealt{bellazzini+12, fabricius+14, kacharov+14, kimmig+15,
  vandenbosch+16, bellini+17, boberg+17, kamann+18}) seem to suggest
that, when properly surveyed, the majority of GCs shows some
signatures of systemic rotation at intermediate distances from the
center.  In addition, rotation has been found in both intermediate-age
(\citealt{davies+11, mackey+13}) and young massive \citep{henault+12}
clusters.  On the theoretical side, a number of studies predict that
massive star clusters are born with significant amounts of rotation,
that is gradually dissipated via the effects of angular moment
transport and loss due to the effects two-body relaxation and star
escape (see e.g., \citealt{longaretti+96, einsel+99, fiestas+06,
  ernst+07, varri+12, vesperini+14, hong+13, mapelli17,
  tiongco+17}). The recent N-body simulations of \citet{tiongco+17}
discuss the long-term evolution of GC rotational properties after an
initial violent relaxation phase (Vesperini et al. 2014) and show that
at the end of that epoch the radial profile of the cluster rotation
velocity displays a well-defined peak at a few half-mass radii in all
the explored models. The combined effect of angular momentum transport
and angular momentum loss due to the escape of stars leads to a
progressive decline in the magnitude of the peak of the rotation curve
with time (see their Fig. 4). The peak is initially located at a
few $r_h$, and it then moves slightly inward over time, but remains
essentially located in a region between 0.5 and 2.5 $r_h$ for most of the
cluster's evolution (see their Figure 6). According to
\citet{tiongco+17} the amplitude of the rotation peak is expected to
decrease as function of time by one order of magnitude, from typical
values of $\sim 0.5 \sigma_0$, down to $\sim 0.05 \sigma_0$ in the
most evolved systems (see their Figure 7).  Although a detailed
comparison between simulations and observations is beyond the scope of
this study, we find a general overall agreement with the predictions
of \citet{tiongco+17}, both in terms of the ratio between $A_{\rm
  rot}$ and $\sigma_0$ (which ranges between 0.1 and 0.3 in our
candidate rotators) and for what concerns the radial location of the
rotation peak (which is found within a few $r_h$ in our observations;
see Table \ref{tab_vrot}).

For the candidate rotators, the rotation amplitudes are of the
order of $\sim 1$-2 km s$^{-1}$ (see Table \ref{tab_vrot}). In
principle, these should be taken into account for the determination of
the true stellar VD that is defined as the square root of
$\sigma^2_P(r) -A^2_{\rm rot}(r)$, where $\sigma_P(r)$ is the observed
root mean square of the RV distribution determined in Section
\ref{sec_vdisp} (see Table \ref{tab_vdisp}). In practice, however, the
ratio between $A_{\rm rot}(r)$ and $\sigma_P(r)$ in the radial bin
where the maximum rotation signal is found is always of the order of
$\sim 0.3-0.4$ and the resulting value of the true VD coincides within
the errors with the measured values of $\sigma_P(r)$. Hence, we can
safely adopt $\sigma_P(r)$ as the true VD profiles of each cluster,
with no real need of corrections for ordered rotation.

\begin{figure}
\includegraphics[width=15cm]{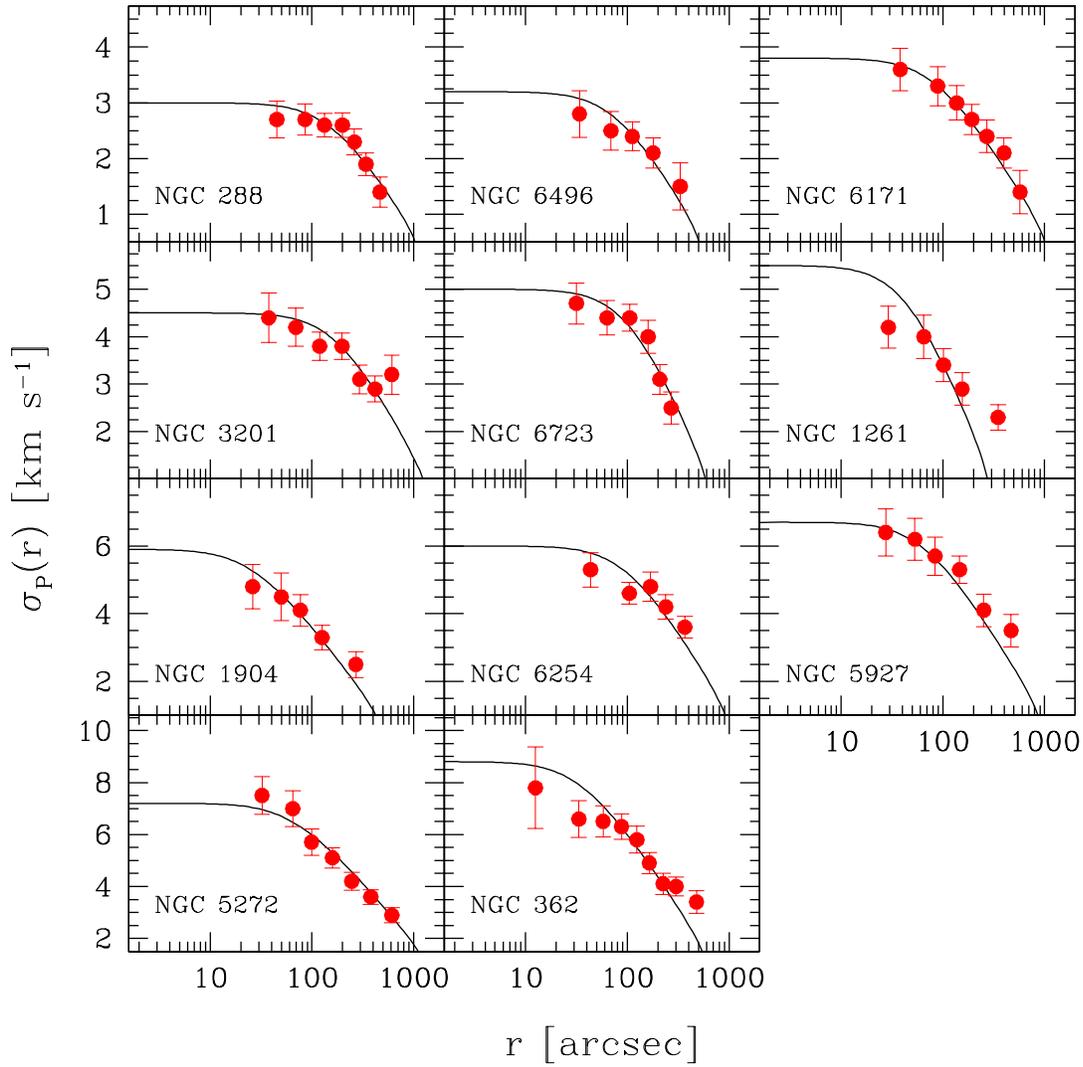}
\caption{Projected velocity dispersion profiles for the program
  clusters as determined from the RV of individual stars surveyed with
  KMOS+FLAMES observations (red filled circles). The solid lines
  correspond to the projected VD profiles of the King models that best
  fit the observed density/surface brightness distributions (see
  Sect. \ref{sec_vdisp}).}
\label{vdisp}
\end{figure}

%-----> VD profile
As shown in Figure \ref{vdisp}, the derived VD profiles sample a
significant radial fraction of each cluster, covering from 3 up to 20
half-mass radii.  As expected for ``well-behaved'' GCs, the stellar VD
profile declines outward.  Indeed, the \citet{king66} models that best
fit the observed density/luminosity distribution of each cluster also
reproduce the projected VD profiles reasonably well (see the solid
lines in Figure \ref{vdisp}). The adopted model parameters are listed
in Table \ref{tab_struct}.  For NGC 6496 the values quoted in
\citet{harris96} and in \citet{McLvdM05} seem to be affected by some
problem (for instance, the half-mass radius almost coincides with the
core radius in the former, while the dimensionless central potential
is $W_0=0.3^{+4.3}_{-0.0}$ in the latter). We therefore used the
photometric data of the ACS Survey of Galactic Globular Clusters
\citep{sarajedini+07} and 2MASS \citep{Skrutskie+06} to obtain a new
determination of the projected density distribution of NGC 6496 from
resolved star counts. The structural parameters of the King model that
best fits the resulting profile are $W_0=5.7$ (corresponding to
concentration $c=1.18$), core radius $r_c=35.6\arcsec$, and half-mass
radius $r_h=93.6\arcsec$. As shown in Figure \ref{vdisp} this model
nicely reproduces also the observed VD profile. Particularly
interesting is the case of NGC 288, for which we detect a clear
decline beyond $r\sim 200\arcsec$ (corresponding roughly to 9 pc for
the distance quoted in \citealp{ferraro+99}; see Table
\ref{tab_struct}).  This is not in agreement with the results of
\citet{hernandez+17}, who find that the VD profile of this cluster
flattens at $r\sim 8$-10 pc and stays constant at $\sigma_P\simeq 2.0$
km/s over the whole radial range sampled. It worth noticing, however,
that the VD curve of \citet{hernandez+17} is obtained from 148 stars
with cluster-centric distances $r\le 16$ pc, while our results is
based on a sample of more than 400 members observed out to $\sim 30$
pc. On the other hand, the observed declining shape of the VD profile
is agreement with the results of \citet{lane+10} and
\citet{kimmig+15}, and it is well matched by the King model that best
fits the star density distribution of the cluster.

%-----> sig0
The King model profiles shown in Figure \ref{vdisp} provide
zeroth-order estimates of the central VD ($\sigma_0$) in each cluster.
These are listed in Table \ref{tab_vsys_sig0}, together with the
values quoted in the Harris catalog for comparison. The two sets of
values are in good agreement for all the GCs in common, with the
exceptions of NGC 362 and NGC 5272 (M3) for which we find
significantly larger central VDs. Our results, however, are in good
agreement with those recently determined by \citet{kimmig+15} and
\citet{kamann+18}, while the values quoted in the Harris catalog
derive from early determinations based on much smaller samples of
spectra.  Within the uncertainties, our central VDs also agree with
those of \citet{lane+10} and \citet{bellazzini+12} for the clusters in
common.  For NGC 5927, instead, we find a lower value ($6.7 \pm 0.7$
km s$^{-1}$) with respect to \citet{lardo+15}, who quote
$\sigma_0=11.0\pm2.0$ km s$^{-1}$. This can be explained by noticing
that the latter value has been derived from the central extrapolation
of a poorly sampled VD profile (79 stars in total, with just $\sim 5$
objects within $60\arcsec$ from the center), while we have 534 members
in total, of which almost 200 are located ar $r<60\arcsec$.

%-----> mass
Under the assumption that the program clusters are well represented by
single-mass, spherical, isotropic and non-rotating \citet{king66}
models, we can use the derived values of the central VD to estimate
the total mass of each system. For this purpose we use equation (3) in
\citet{majewski+03}, where the parameters $\mu$ and $\beta$ have been
determined, respectively, by following \citet{djorgovski93} and by
assuming $\beta=1/\sigma_0$ (as appropriate for models with $W_0>5$;
see the discussion in \citealp{richstone+86}). The resulting masses
are listed in Table \ref{tab_mass}.  They agree within a factor of
$\sim 2$ with the values quoted in the literature (see Table
\ref{tab_mass} and also \citealp{lane+10, zocchi+12, kimmig+15}). This
is well acceptable if one takes into account all the uncertainties and
the fact that the various estimates have been obtained through
different methods (for instance, \citealp{McLvdM05} use total
luminosities and population-synthesis $V$-band mass-to-light ratio
ratios, while \citealp{baum17} uses multi-mass N-body simulations).

\section{Summary and Conclusions}
\label{sec_conclu}
This work is part of the MIKis survey, a project aimed at providing
the line-of-sight kinematic information along the entire radial
extension of a selected sample of 30 GGCs.  To this purpose, we
exploit large and homogeneous datasets of RVs, measured from
medium-high resolution spectra of individual stars acquired through
the combined use of three ESO-VLT spectrographs: diffraction-limited
IF observations with SINFONI for the innermost cluster regions, and
KMOS and FLAMES data for the intermediate and external radial ranges,
respectively.  Here we presented the first results obtained for 11
GGCs in the survey. We provide the first determinations of the VD
profile and systemic rotation information for NGC 1261 and NGC
6496. For the latter, we also present updated structural parameters,
obtained from the construction of a new density profile from resolved
star counts and its \citet{king66} model best-fit.  All the observed
VD profiles decline outward and, at a first approximation, they are
reproduced by the same King model that best-fits the
density/luminosity distribution. We found evidence of rotation at 1-2
$r_h$ in the majority of the surveyed clusters.  Together with other
findings in the literature, this suggests that possibly most (if not
all) GGCs display some degree of internal rotation, which might be the
remnant of a much larger amount of ordered motions imprinted at birth
and then gradually dissipated via two-body relaxation
\citep[e.g.,][]{fiestas+06, tiongco+17}.  Hence, particular care
should be devoted to explore the rotational properties of GGCs, since
the detection of even weak signals is not an indication of the lack of
importance of rotation in these systems, but it possibly represents
the observational evidence that most of the clusters were born with
significant amounts of ordered motions \citep{tiongco+17}.

\begin{figure}
\includegraphics[width=15cm]{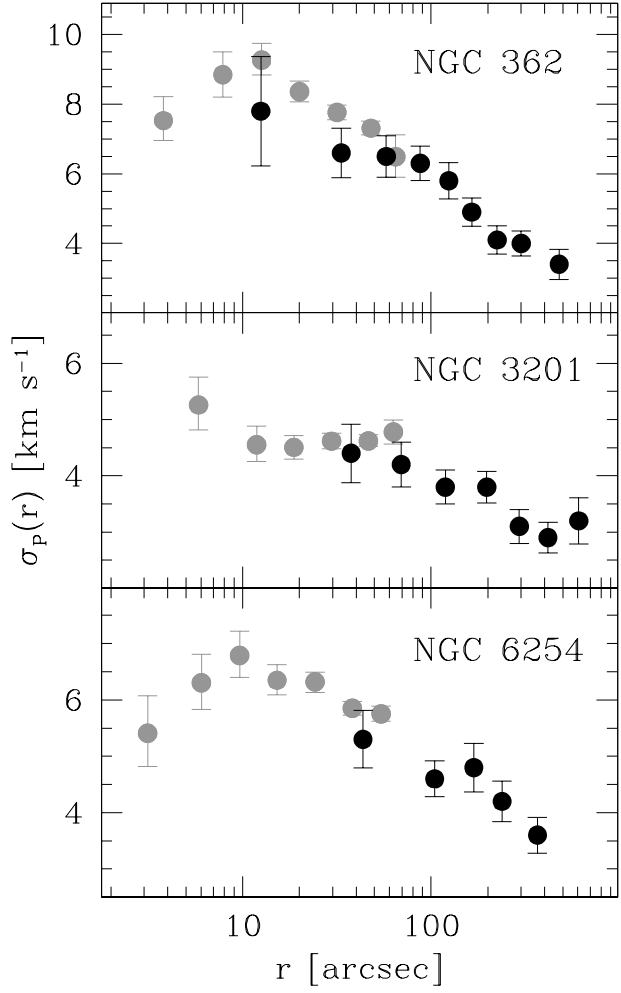}
\caption{Projected VD profile of the three clusters in common with
  \citet{kamann+18}.  Filled black circles are from this work, while
  the grey circles are the results of \citet{kamann+18}.}
\label{fig_confro_kam}
\end{figure}

\begin{figure}
\includegraphics[width=15cm]{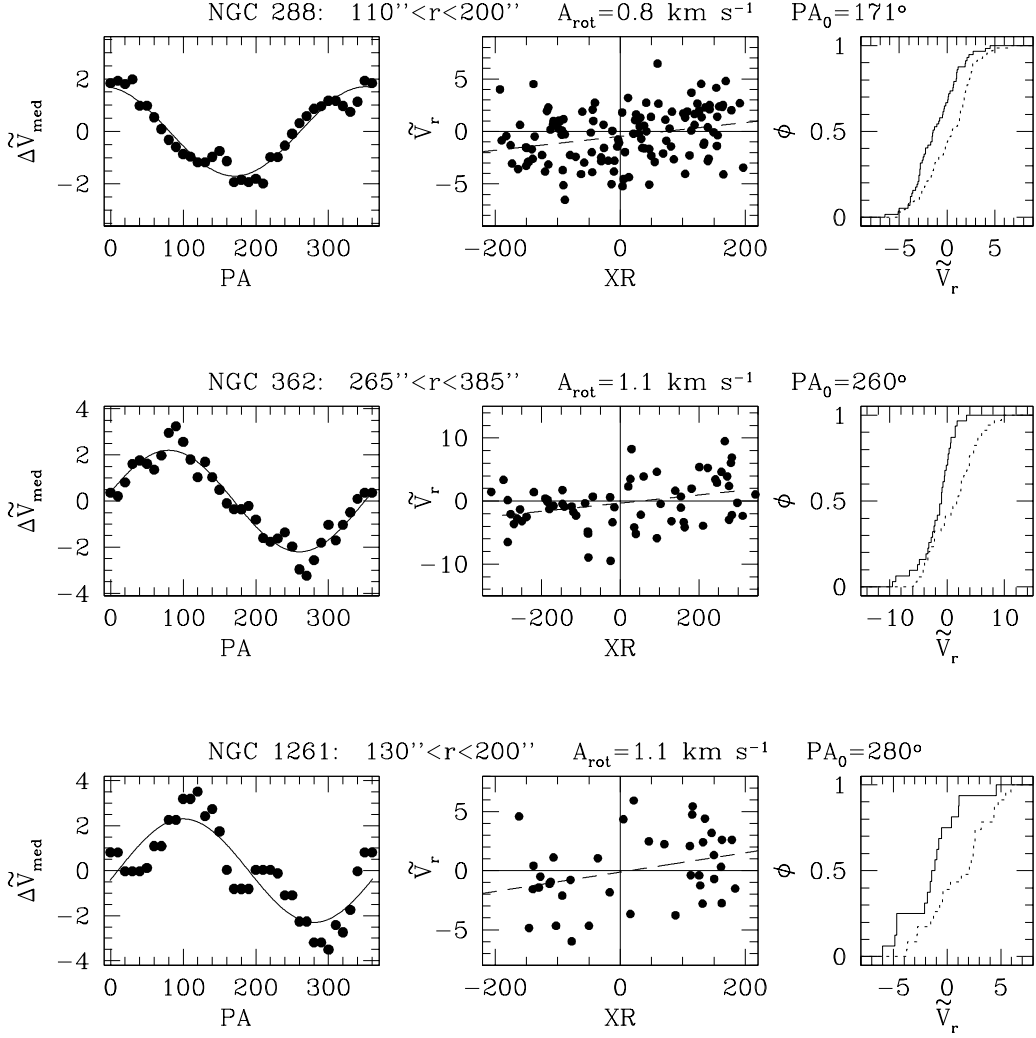}
\caption{Diagnostic diagrams of the rotation signature detected in NGC
  288, NGC 362 and NGC 1261 (top, middle, and bottom rows,
  respectively). For each system, the {\it left panel} shows the
  difference between the median RVs on each side of the cluster with
  respect to a line passing through the center with a given position
  angle (PA), as a function of PA itself (see Section
  \ref{sec_vrot}). The continuos line is the sine function that best
  fits the observed pattern. The best-fit rotation amplitude and
  position angle (A$_{\rm rot}$ and PA$_0$, respectively) are labelled
  above the panels, together with the considered radial range.  The
  {\it central panel} shows the distribution of the measured RVs
  ($\widetilde V_r =V_r-V_{\rm sys}$) as a function of the projected
  distances from the rotation axis (XR) in arcseconds. The dashed line
  is the least square fit to the data.  The {\it right panel} shows
  the cumulative RV distributions for the sample of stars with XR$< 0$
  (solid line) and for that with XR$>0$ (dotted line).}
\label{fig_vrot1}
\end{figure}

\begin{figure}
\includegraphics[width=15cm]{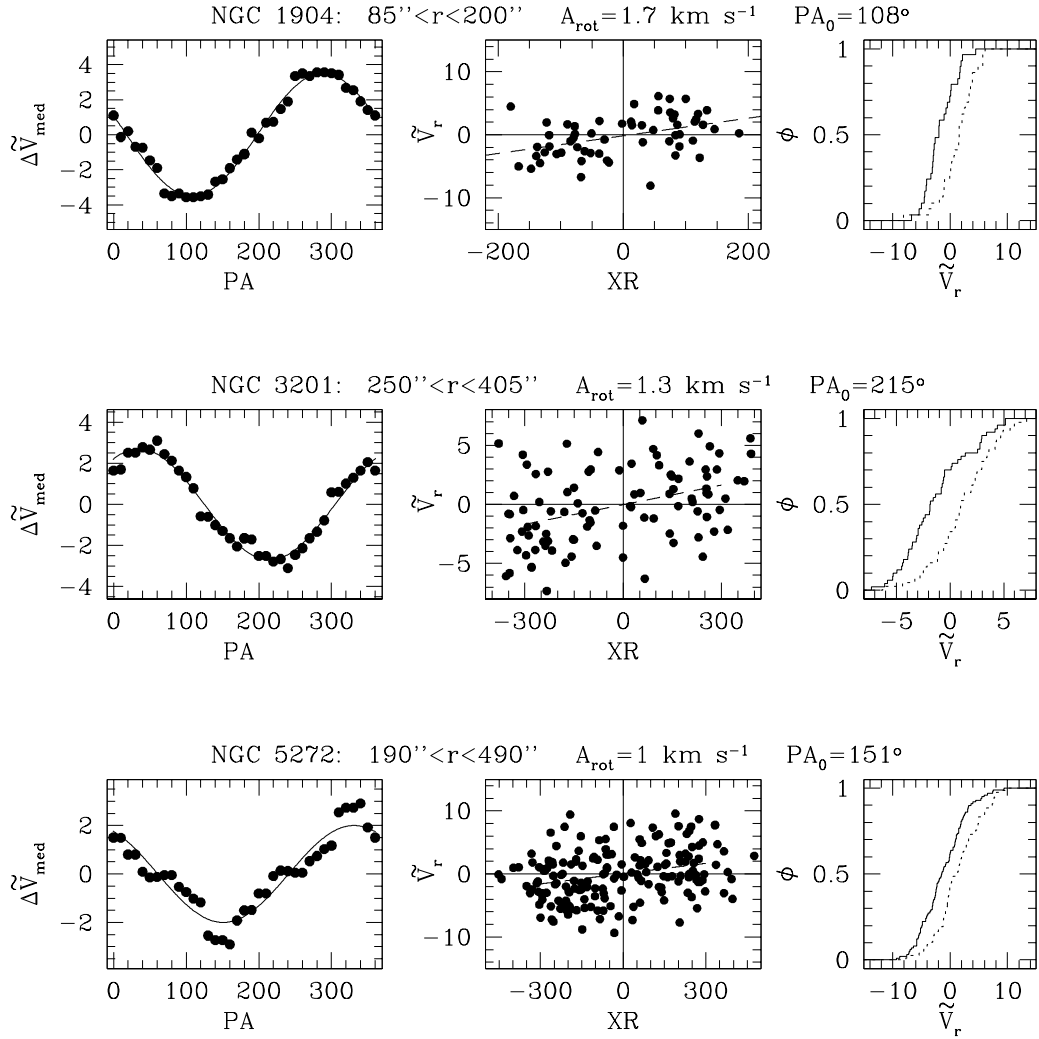}
\caption{As in Figure \ref{fig_vrot1}, but for NGC 1904, NGC 3201, NGC 5272.}
\label{fig_vrot2}
\end{figure}

\begin{figure}
\includegraphics[width=15cm]{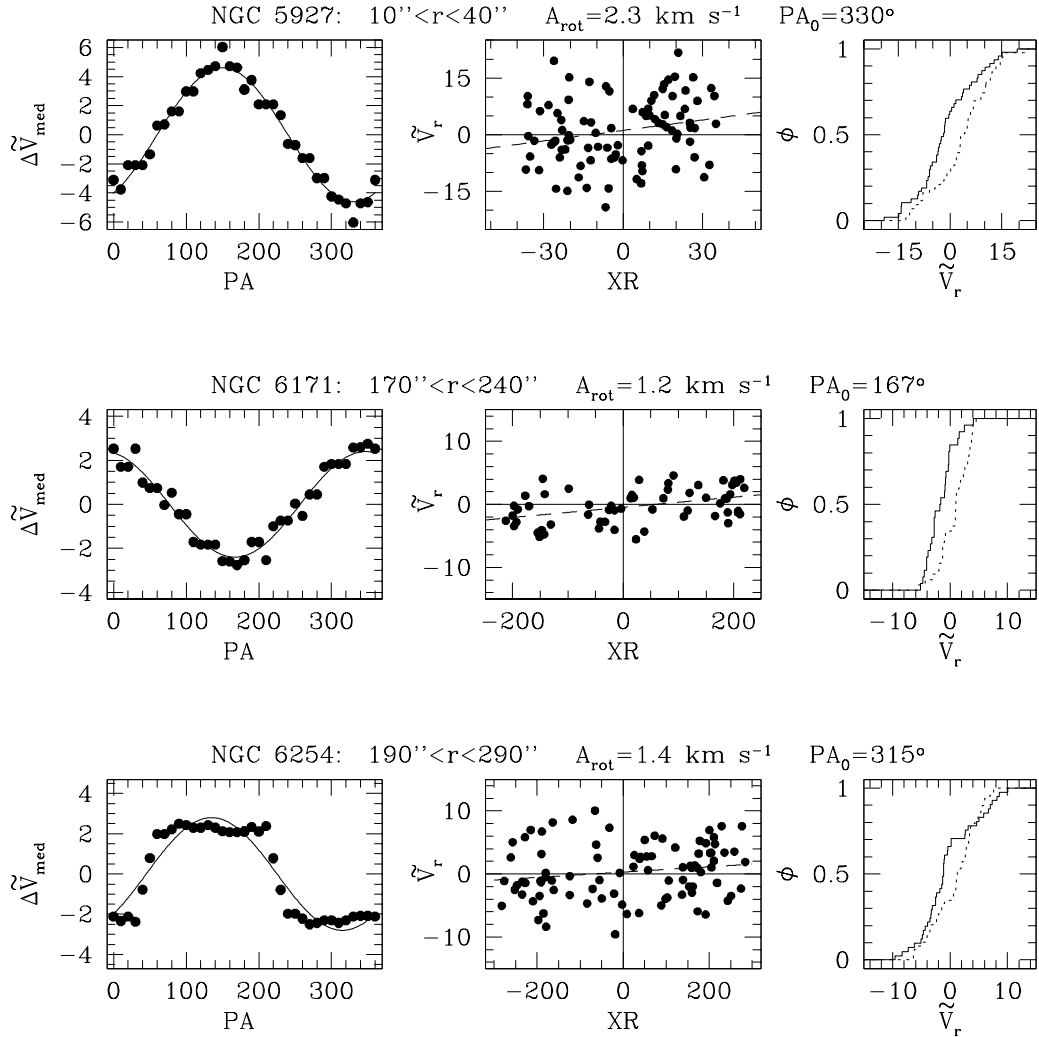}
\caption{As in Figure \ref{fig_vrot1}, but for NGC 5927, NGC 6171, NGC 6254.}
\label{fig_vrot3}
\end{figure}

\begin{figure}
\includegraphics[width=15cm]{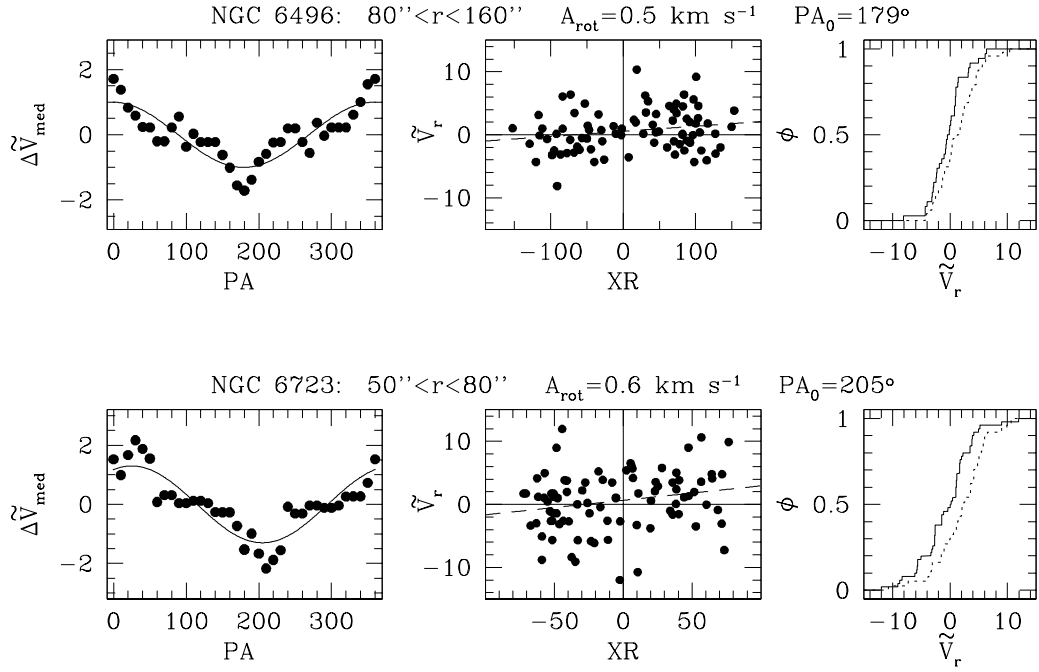}
\caption{As in Figure \ref{fig_vrot1}, but for NGC 6496 and NGC 6723.}
\label{fig_vrot4}
\end{figure}

The data presented here can be now complemented with observations of
the central regions in order to obtain a full radial coverage of each
cluster. A nice example based on MUSE data (from \citealp{kamann+18})
is shown in Figure \ref{fig_confro_kam}, but an even a better spatial
resolution to explore the innermost region of high density GCs can be
reached with the enhanced version of MUSE (WFM-AO, which operates
under super-seeing conditions down to $FWHM\sim 0.4\arcsec$), or by
using SINFONI (see \citealp{lanzoni+13}).  In a series of future
papers we will present the detailed kinematic study of other specific
GGCs, performed by exploiting the powerful multi-instrument dataset
acquired within the MIKis survey.  The kinematic information along the
line-of-sight thus obtained may then be combined with star density
profiles (now feasible for most GCs, see \citealt{miocchi+13}) and,
for selected clusters, also with the structural and kinematic maps on
the plane of the sky as obtained from new-generation astrometric data
in the central and outer regions, from HST and Gaia,
respectively. Such a rich view of the phase space of these systems
will enable a complete dynamical interpretation, by means of
state-of-the-art equilibrium and evolutionary models (e.g.,
\citealp{varri+12} and \citealp{wang+16}, respectively).

This first exploration of our rich kinematic survey already proves to
be of outstanding value, in at least two respects. On the one hand, it
coronates the mounting empirical evidence (see references in the
previous section) that a new level of sophistication may now be
attained in the characterization of the velocity space of Galactic
GCs, unveiling an unexpected degree of kinematic richness, which makes
them refreshingly novel `phase space laboratories'.  On the other
hand, it also provides the motivation and the opportunity to deepen
the theoretical exploration of a number of fundamental aspects of
collisional stellar dynamics, such as the role of angular momentum,
orbital anisotropy and their interplay with the external tidal
field. A full theoretical understanding and a detailed observational
investigation of the complete velocity space of GCs are essential
steps towards an appropriate dynamical interpretation of a number of
outstanding puzzles of this class of stellar systems, especially
related to their elusive stellar populations and putative
intermediate-mass black holes.

\acknowledgements{We thank the anonymous referee for useful comments
  that contributed to improve the presentation of the paper. FRF
  acknowledges the ESO Visitor Programme for the support and the warm
  hospitality at the ESO Headquarter in Garching (Germany) during the
  period when part of this work was performed. ALV acknowledges
  support from a Marie Sklodowska-Curie Fellowship (MSCA-IF-EFRI
  658088).}

\newpage
\begin{table}[h!]
\begin{center}
\begin{tabular}{l l | l l l| l l l }
\hline
CLUSTER   &         &               & FLAMES        &              &              & KMOS         &                              \\	
\hline	      
NGC 288   &         & 2$\times$1770 s& 1$\times$2670 s&                &               &  2$\times$60 &  1$\times$100 s                  \\
NGC 362   &         & 1$\times$900 s & 2$\times$1770 s&                &  7$\times$30 s&              &  4$\times$100 s                 \\
NGC 1261  &         & 2$\times$1770 s& 2$\times$2670 s&                &               &              &  5$\times$100 s                 \\
NGC 1904  & M79     & 2$\times$2670 s&                &                &  2$\times$30 s&              &  2$\times$100 s                 \\
NGC 3201  &         & 1$\times$900 s & 3$\times$1770 s&                &  6$\times$30 s&  2$\times$60 &  2$\times$100 s                 \\
NGC 5272  & M3      & 1$\times$900 s & 3$\times$1800 s& 2$\times$2670 s&  8$\times$30 s&              &  2$\times$100 s                 \\
NGC 5927  &         & 4$\times$1800 s& 1$\times$2670 s&                &  4$\times$30 s&              &  5$\times$100 s                 \\
NGC 6171  &         & 2$\times$900 s & 1$\times$1800 s& 1$\times$2700 s&               &  5$\times$60 &  2$\times$100 s                 \\
NGC 6254  & M10     & 1$\times$1200 s& 1$\times$1331 s& 2$\times$1800 s&  4$\times$30 s&              &  2$\times$100 s                 \\
NGC 6496  &         & 3$\times$1800 s& 1$\times$2700 s& 1$\times$5400 s&               &              &  5$\times$300 s  \\
NGC 6723  &         & 3$\times$900 s & 3$\times$1800 s& 1$\times$2700 s&  5$\times$30 s&              &  4$\times$100 s                 \\
\hline
\end{tabular}
\end{center}
\caption{Number and duration of the exposures secured for each cluster
  within the MIKiS survey.}
\label{tab_obs}
\end{table}

\newpage
\begin{table}[h!]
\begin{center}
\begin{tabular}{l l l l c}
\hline
Cluster   &        & Program ID      &                        & Grating   \\                    
\hline
NGC 288   &        &  074.A-0508  & (PI Drinkwater)	   & LR2-LR4	  \\
          &        &  073.D-0211  & (PI Carretta)	   & HR11-HR13	  \\
          &        &  075.D-0043  & (PI Carraro)	   & HR9	  \\
          &        &  087.D-0276  & (PI D'Orazi)	   & HR15-HR19	  \\
          &        &  088.B-0403  & (PI Lucatello)	   & HR9	  \\
          &        &  193.D-0232  & (PI Ferraro)	   & HR21	  \\
\hline
NGC 362   &        &  083.D-0208  & (PI Carretta)	   & HR11-HR13	  \\
          &        &  087.D-0276  & (PI D'Orazi)	   & HR15N	  \\
          &        &  088.D-0026  & (PI Mc Donald)	   & HR14-HR15	  \\
          &        &  188.B-3002  & (PI Gilmore)	   & HR10-HR21	  \\
          &        &  193.D-0232  & (PI Ferraro)	   & HR21	  \\
\hline
NGC 1261  &        &  193.D-0232  & (PI Ferraro)	   & HR21	  \\
          &        &  188.B-3002  & (PI Gilmore)     & HR10-HR21 \\
          &        &  193.B-0936  & (PI Gilmore)     & HR10-HR21 \\
\hline
NGC 1904  & M79    &  072.D-0507  & (PI Carretta)	   & HR11-HR13	  \\
          &        &  085.D-0205  & (PI Carretta)	   & HR21	  \\
          &        &  193.B.0936  & (PI Gilmore)	   & HR10-HR21	  \\
          &        &  193.D-0232  & (PI Ferraro)	   & HR21	  \\
\hline
NGC 3201  &        &  171.B-0520  & (PI Gilmore)	   & LR8	  \\
          &        &  073.D-0211  & (PI Carretta)	   & HR11-HR13	  \\
          &        &  087.D-0276  & (PI D'Orazi)	   & HR15-HR19	  \\
          &        &  088.B-0403  & (PI Lucatello)	   & HR9	  \\
          &        &  193.D-0232  & (PI Ferraro)	   & HR21	  \\
\hline
NGC 5272  & M3     &  093.D-0536  & (PI Contreras Ramos)   & HR12	      \\
          &        &  193.D-0232  & (PI Ferraro)	   & HR21	  \\
\hline
\hline
\end{tabular}
\end{center}
\caption{Summary of the FLAMES dataset sets used to derive the
  internal kinematics of the target GCs.  For each system the proposal
  ID, the PI and the used grating are listed. The datasets acquired
  within the MIKiS survey correspond to Program ID 193.D-0232, while
  the others have been retrieved from the ESO archive.}
\label{tab_datasets}
\end{table}

\newpage
\begin{table}[h!]%%\ContinuedFloat
\begin{center}
\begin{tabular}{l l l l c}
\hline
CLUSTER   &        & PROGRAM      &                        & GRATING   \\                    
\hline
NGC 5927  &        &  079.B-0721  & (PI Feltzing)	   & HR13	  \\
          &        &  188.B-3002  & (PI Gilmore)	   & HR10-HR21	  \\
          &        &  193.D-0232  & (PI Ferraro)	   & HR21	  \\    
\hline
NGC 6171  &        &  073.D-0211  & (PI Carretta)	   & HR13	  \\ 
          &        &  193.D-0232  & (PI Ferraro)	   & HR21	  \\
          &        &  071.D-0311  & (PI Scarpa)     & HR9       \\ 
\hline
NGC 6254  & M10    &  073.D-0211  & (PI Carretta)	   & HR11-HR13	  \\ 
          &        &  193.B-0232  & (PI Ferraro)	   & HR21	  \\
\hline
NGC 6496  &        &  193.D-0232  & (PI Ferraro)	   & HR21	  \\
\hline
NGC 6723  &        &  087.D-0230  & (PI Gratton)	   & HR12-HR19	  \\ 
          &        &  193.D-0232  & (PI Ferraro)	   & HR21	  \\
\hline
\hline
\end{tabular}
\end{center}
\caption*{}{Table \ref{tab_datasets} (continued)}
\end{table}

\newpage
\begin{table}[h!]
\begin{center}
\begin{tabular}{lcccc}
\hline
Cluster  & $N_{\rm obs}$ & $r_{\rm min}$ & $r_{\rm max}$ & $N_{\rm memb}$ \\
         &              &  [arcsec]    &  [arcsec]   & \\
\hline
NGC  288 &  538  &  1.0  & 853 & 419 \\
NGC  362 &  717  &  1.0  & 815 & 543 \\ 
NGC 1261 &  320  &  1.2  & 728 & 299 \\
NGC 1904 &  235  &  3.8  & 774 & 173 \\
NGC 3201 &  587  &  4.7  & 749 & 454 \\
NGC 5272 &  628  &  1.9  & 750 & 577 \\
NGC 5927 &  851  &  5.2  & 784 & 534 \\
NGC 6171 &  482  &  3.0  & 746 & 319 \\
NGC 6254 &  565  &  9.8  & 775 & 415 \\
NGC 6496 &  656  &  8.6  & 753 & 234 \\
NGC 6723 &  696  &  2.1  & 731 & 487 \\
\hline
\end{tabular}
\end{center}
\caption{Total number of observed stars ($N_{\rm obs}$), minimum and
  maximum distances fron the cluster center sampled by the RV datasets
  ($r_{\rm min}$ and $r_{\rm max}$, respectively), and number of
  member stars used for the determination of the VD profile and the
  search for systemic rotation ($N_{\rm memb}$).}
\label{tab_nstar}
\end{table}

%\newpage
\begin{table}[h!]
\begin{center}
%\begin{tabular}{lcccc}
\begin{tabular}{lrrrr}
\hline Cluster & $V_{\rm sys}$~~~ & $\sigma_0$~~~   & $V_{\rm sys, H}$~~~ & $\sigma_{0, H}$ ~~~\\
               & [km s$^{-1}$]& [km s$^{-1}$] & [km s$^{-1}$]   & [km s$^{-1}$] \\
\hline
NGC  288 &  $ -44.6\pm 0.4$  &  $3.0\pm 0.3$  & $ -45.4\pm 0.2$  & $2.9\pm 0.3$ \\
NGC  362 &  $ 222.5\pm 0.4$  &  $8.8\pm 1.6$  & $ 223.5\pm 0.5$  & $6.4\pm 0.3$ \\ 
NGC 1261 &  $  71.6\pm 0.6$  &  $5.5\pm 0.4$  & $  68.2\pm 4.6$  & $-  $\\
NGC 1904 &  $ 205.4\pm 0.6$  &  $5.9\pm 0.7$  & $ 205.8\pm 0.4$  & $5.3\pm 0.4$ \\
NGC 3201 &  $ 494.5\pm 0.4$  &  $4.5\pm 0.5$  & $ 494.0\pm 0.2$  & $5.0\pm 0.2$ \\
NGC 5272 &  $-147.2\pm 0.4$  &  $7.2\pm 0.7$  & $-147.6\pm 0.2$  & $5.5\pm 0.3$ \\
NGC 5927 &  $-104.6\pm 0.4$  &  $6.7\pm 0.7$  & $-107.5\pm 0.9$  & $- $~~~ \\
NGC 6171 &  $ -34.4\pm 0.5$  &  $3.8\pm 0.4$  & $ -34.1\pm 0.3$  & $4.1\pm 0.3$ \\
NGC 6254 &  $  75.8\pm 0.4$  &  $6.0\pm 0.5$  & $  75.2\pm 0.7$  & $6.6\pm 0.8$ \\
NGC 6496 &  $-134.6\pm 0.7$  &  $3.2\pm 0.4$  & $-112.7\pm 5.7$  & $- $~~~ \\
NGC 6723 &  $ -95.3\pm 0.4$  &  $5.0\pm 0.4$  & $ -94.5\pm 3.6$  & $- $~~~ \\
\hline
\end{tabular}
\end{center}
\caption{Systemic velocity ($V_{\rm sys}$) and central velocity
  dispersion ($\sigma_0$) of the program clusters, as determined in
  the present work (columns 2 and 3, respectively), and as quoted in
  the \citet{harris96} catalog (last two columns).}
\label{tab_vsys_sig0}
\end{table}

\newpage
\begin{table}[h!]
\begin{center}
\begin{tabular}{cccccc}
\hline
\multicolumn{6}{c}{\bf{NGC 288}}\\
\hline
  $r_i$   &  $r_e$    &   $r_m$  &  N  & $\sigma_p$  & $err_\sigma$ \\
  arcsec  &  arcsec   &   arcsec &     & km s$^{-1}$ & km s$^{-1}$  \\
  \hline
  15.00 &  65.00 &  45.12 &  39  &  2.70  & 0.33 \\
  65.00 & 105.00 &  85.49 &  64  &  2.70  & 0.28 \\
 105.00 & 165.00 & 132.62 &  86  &  2.60  & 0.21 \\
 165.00 & 235.00 & 199.32 &  83  &  2.60  & 0.22 \\
 235.00 & 295.00 & 261.11 &  57  &  2.30  & 0.23 \\
 295.00 & 395.00 & 339.94 &  62  &  1.90  & 0.20 \\
 395.00 & 645.00 & 465.92 &  28  &  1.40  & 0.27 \\
\hline
\hline
\multicolumn{6}{c}{\bf{NGC 362}}\\
\hline
  $r_i$   &  $r_e$    &   $r_m$  &  N  & $\sigma_p$  & $err_\sigma$ \\
  arcsec  &  arcsec   &   arcsec &     & km s$^{-1}$ & km s$^{-1}$  \\
  \hline
   3.00 &  20.00 &  12.49  &  33  &  7.80  & 1.57 \\
  20.00 &  45.00 &  33.37  &  70  &  6.60  & 0.71 \\
  45.00 &  70.00 &  57.79  &  65  &  6.50  & 0.60 \\
  70.00 & 110.00 &  87.53  &  88  &  6.30  & 0.49 \\
 110.00 & 140.00 & 124.19  &  64  &  5.80  & 0.52 \\
 140.00 & 200.00 & 164.43  &  73  &  4.90  & 0.41 \\
 200.00 & 260.00 & 223.99  &  50  &  4.10  & 0.41 \\
 260.00 & 350.00 & 300.89  &  65  &  4.00  & 0.36 \\
 350.00 & 700.00 & 478.30  &  35  &  3.40  & 0.43 \\
\hline
\hline
\multicolumn{6}{c}{\bf{NGC 1261}}\\
\hline
  $r_i$   &  $r_e$    &   $r_m$  &  N  & $\sigma_p$  & $err_\sigma$ \\
  arcsec  &  arcsec   &   arcsec &     & km s$^{-1}$ & km s$^{-1}$  \\
  \hline
   1.00  &  50.00 &  29.15 & 107 &  4.20 & 0.44 \\
  50.00  &  80.00 &  65.11 &  59 &  4.00 & 0.46 \\
  80.00  & 130.00 & 101.61 &  57 &  3.40 & 0.35 \\
 130.00  & 200.00 & 155.01 &  39 &  2.90 & 0.34 \\
 200.00  & 800.00 & 349.32 &  37 &  2.30 & 0.27 \\
\hline
\hline
\multicolumn{6}{c}{\bf{NGC 1904}}\\
\hline
  $r_i$   &  $r_e$    &   $r_m$  &  $N$  & $\sigma_p$  & $err_\sigma$ \\
  arcsec  &  arcsec   &   arcsec &     & km s$^{-1}$ & km s$^{-1}$  \\
  \hline
   5.00 &  40.00 &   26.13 &  39 &  4.80 & 0.65 \\
  40.00 &  60.00 &   49.94 &  28 &  4.50 & 0.70 \\
  60.00 & 100.00 &   76.89 &  42 &  4.10 & 0.47 \\ 
 100.00 & 170.00 &  125.62 &  41 &  3.30 & 0.36 \\
 170.00 & 600.00 &  270.13 &  23 &  2.50 & 0.38 \\
\hline
\hline
\end{tabular}
\end{center}
\caption{VD profiles of the program clusters: internal and external
  radius of each radial bin ($r_i$ and $r_e$, respectively), average
  cluster-centric distance of the member stars in the bin ($r_m$),
  number of stars in the bin ($N$), measured VD and its uncertainty in
  the bin ($\sigma_p$ and $err_\sigma$, respectively).}
\label{tab_vdisp}
\end{table}

\newpage
\begin{table}[h!]%\ContinuedFloat
\begin{center}
\begin{tabular}{cccccc}
\hline
\multicolumn{6}{c}{\bf{NGC 3201}}\\
\hline
  $r_i$   &  $r_e$    &   $r_m$  &  N  & $\sigma_p$  & $err_\sigma$ \\
  arcsec  &  arcsec   &   arcsec &     & km s$^{-1}$ & km s$^{-1}$  \\
  \hline
  20.00 &  50.00 &  37.62  &  39  & 4.40 & 0.52 \\
  50.00 &  90.00 &  69.42  &  59  & 4.20 & 0.40 \\
  90.00 & 150.00 & 119.05  &  84  & 3.80 & 0.30 \\
 150.00 & 250.00 & 197.37  &  99  & 3.80 & 0.28 \\
 250.00 & 350.00 & 293.92  &  64  & 3.10 & 0.30 \\
 350.00 & 500.00 & 417.33  &  66  & 2.90 & 0.27 \\
 500.00 & 800.00 & 608.31  &  43  & 3.20 & 0.41 \\
\hline
\hline
\multicolumn{6}{c}{\bf{NGC 5272}}\\
\hline
  $r_i$   &  $r_e$    &   $r_m$  &  N  & $\sigma_p$  & $err_\sigma$ \\
  arcsec  &  arcsec   &   arcsec &     & km s$^{-1}$ & km s$^{-1}$  \\
  \hline
   1.00 &  50.00 &   32.31 &   99 &  7.50  & 0.72 \\
  50.00 &  80.00 &   64.86 &   69 &  7.00  & 0.69 \\
  80.00 & 120.00 &   99.09 &   72 &  5.70  & 0.51 \\
 120.00 & 200.00 &  158.67 &  110 &  5.10  & 0.39 \\
 200.00 & 300.00 &  246.19 &   83 &  4.20  & 0.35 \\
 300.00 & 500.00 &  378.94 &   91 &  3.60  & 0.28 \\
 500.00 & 770.00 &  610.01 &   53 &  2.90  & 0.29 \\
\hline
\hline
\multicolumn{6}{c}{\bf{NGC 5927}}\\
\hline
  $r_i$   &  $r_e$    &   $r_m$  &  N  & $\sigma_p$  & $err_\sigma$ \\
  arcsec  &  arcsec   &   arcsec &     & km s$^{-1}$ & km s$^{-1}$  \\
  \hline
   1.00 &  40.00 &  27.49 &   96 &  6.40 &  0.70 \\
  40.00 &  70.00 &  53.18 &  109 &  6.20 &  0.62 \\
  70.00 & 100.00 &  83.96 &   67 &  5.70 &  0.56 \\
 100.00 & 200.00 & 146.35 &  129 &  5.30 &  0.40 \\
 200.00 & 300.00 & 252.78 &   56 &  4.10 &  0.48 \\
 300.00 & 800.00 & 469.22 &   56 &  3.50 &  0.48 \\
\hline
\hline
\multicolumn{6}{c}{\bf{NGC 6171}}\\
\hline
  $r_i$   &  $r_e$    &   $r_m$  &  N  & $\sigma_p$  & $err_\sigma$ \\
  arcsec  &  arcsec   &   arcsec &     & km s$^{-1}$ & km s$^{-1}$  \\
  \hline
   2.00 &  65.00 &  38.12 &  72 &  3.60 & 0.38 \\
  65.00 & 110.00 &  89.06 &  54 &  3.30 & 0.35 \\
 110.00 & 160.00 & 136.78 &  51 &  3.00 & 0.31 \\
 160.00 & 220.00 & 192.80 &  59 &  2.70 & 0.27 \\
 220.00 & 320.00 & 271.18 &  39 &  2.40 & 0.29 \\
 320.00 & 500.00 & 399.60 &  34 &  2.10 & 0.27 \\
 500.00 & 750.00 & 574.61 &  10 &  1.40 & 0.39 \\
\hline
\hline
\end{tabular}
\end{center}
\caption*{}{Table \ref{tab_vdisp} (continued)}
%\label{tab_vdisp}
\end{table}

\newpage
\begin{table}[h!]%\ContinuedFloat
\begin{center}
\begin{tabular}{cccccc}
\hline
\multicolumn{6}{c}{\bf{NGC 6254}}\\
\hline
  $r_i$   &  $r_e$    &   $r_m$  &  N  & $\sigma_p$  & $err_\sigma$ \\
  arcsec  &  arcsec   &   arcsec &     & km s$^{-1}$ & km s$^{-1}$  \\
  \hline
   1.00 &  70.00 &   43.52 &  85 &  5.30 & 0.51 \\
  70.00 & 140.00 &  104.48 & 114 &  4.60 & 0.32 \\
 140.00 & 200.00 &  168.63 &  73 &  4.80 & 0.43 \\
 200.00 & 280.00 &  238.30 &  73 &  4.20 & 0.36 \\
 280.00 & 550.00 &  367.07 &  70 &  3.60 & 0.32 \\
\hline
\hline
\multicolumn{6}{c}{\bf{NGC 6496}}\\
\hline
  $r_i$   &  $r_e$    &   $r_m$  &  N  & $\sigma_p$  & $err_\sigma$ \\
  arcsec  &  arcsec   &   arcsec &     & km s$^{-1}$ & km s$^{-1}$  \\
  \hline
   8.00 &  50.00 &  33.85 &  62 &  2.80 & 0.42 \\
  50.00 &  85.00 &  68.93 &  60 &  2.50 & 0.35 \\
  85.00 & 140.00 & 111.90 &  62 &  2.40 & 0.26 \\
 140.00 & 250.00 & 178.99 &  37 &  2.10 & 0.27 \\
 250.00 & 450.00 & 330.70 &  13 &  1.50 & 0.42 \\
\hline
\hline
\multicolumn{6}{c}{\bf{NGC 6723}}\\
\hline
  $r_i$   &  $r_e$    &   $r_m$  &  N  & $\sigma_p$  & $err_\sigma$ \\
  arcsec  &  arcsec   &   arcsec &     & km s$^{-1}$ & km s$^{-1}$  \\
  \hline
   5.00 &  50.00 &   31.56 & 112 &  4.70 &  0.43 \\
  50.00 &  80.00 &   63.22 &  87 &  4.40 &  0.36 \\
  80.00 & 140.00 &  105.47 & 130 &  4.40 &  0.28 \\
 140.00 & 180.00 &  159.86 &  71 &  4.00 &  0.35 \\
 180.00 & 240.00 &  208.16 &  58 &  3.10 &  0.31 \\
 240.00 & 320.00 &  269.32 &  29 &  2.50 &  0.34 \\
\hline
\end{tabular}
\end{center}
\caption*{Table \ref{tab_vdisp} (continued)}
%\label{tab_vdisp}
\end{table}

\newpage
\begin{table}[h!]
\begin{center}
%\begin{tabular}{lrrrrrrrrrrr}
\begin{tabular}{lccccccccccc}
\hline
Cluster & $d_{\rm max}$ & $r_i$ & $r_e$ & $r_m$ & $N$ & A$_{\rm rot}$ & PA$_0$ & $P_{\rm KS}$ & $P_{\rm Stud}$ & n-$\sigma_{\rm ML}$ & N$_{\rm bin}$\\
\hline
NGC  288 &  400 & 110 & 200 & 151.20 & 123 & 0.8 & 171 & $1.9 \times 10^{-3}$ & $>99.0$    & 3.8 & 3\\
NGC  362 &  400 & 265 & 385 & 309.20 &  65 & 1.1 & 260 & $9.8 \times 10^{-4}$ & $>99.8$    & 3.5 & 8\\
NGC 1261 &  200 & 130 & 200 & 155.70 &  39 & 1.1 & 280 & $1.0 \times 10^{-2}$ & $>99.0$    & 2.4 & 6\\
NGC 1904 &  200 &  85 & 200 & 124.60 &  58 & 1.7 & 108 & $1.3 \times 10^{-3}$ & $>99.8$    & 5.1 & 3\\
NGC 3201 &  600 & 250 & 405 & 320.40 &  94 & 1.3 & 215 & $9.0 \times 10^{-5}$ & $>99.8$    & 4.3 & 5\\
NGC 5272 &  750 & 190 & 490 & 305.80 & 177 & 1.0 & 151 & $1.5 \times 10^{-4}$ & $>99.8$    & 4.6 & 4\\
NGC 5927 &  100 &  10 &  40 &  28.10 &  90 & 2.3 & 330 & $2.3 \times 10^{-3}$ & $>95.0$    & 2.0 & 3\\
NGC 6171 &  500 & 170 & 240 & 199.60 &  58 & 1.2 & 167 & $8.0 \times 10^{-4}$ & $>99.8$    & 3.8 & 5\\
NGC 6254 &  450 & 190 & 290 & 235.30 &  90 & 1.4 & 315 & $4.2 \times 10^{-3}$ & $<90.0$    & 1.5 & 5\\
NGC 6496 &  200 &  80 & 160 & 116.40 &  84 & 0.5 & 179 & $1.5 \times 10^{-2}$ & $>95.0$    & 3.0 & 4\\
NGC 6723 &  200 &  50 &  80 &  63.40 &  87 & 0.6 & 205 & $6.8 \times 10^{-3}$ & $>99.0$    & 2.5 & 4\\
\hline
\end{tabular}
\end{center}
\caption{Strongest rotation signatures detected in the surveyed
  clusters. For each system, the radial bin where the signal is found
  is labelled in Figures \ref{fig_vrot1}--\ref{fig_vrot4}, while the
  table lists: the maximum distance out to which rotation have been
  studied ($d_{\rm max}$), the internal and external radii of each
  radial bin ($r_i$ and $r_e$, respectively), the mean radius and the
  number of the stars in the bin used to determine the rotation ($r_m$
  and $N$, respectively), the rotation amplitude (A$_{\rm rot}$) and
  the position angle of the rotation axis (PA$_0$) in the bin, the
  Kolmogorov-Smirnov probability that the two samples on each side of
  the rotation axis are drawn from the same parent distribution
  ($P_{\rm KS}$), the t-Student probability that the two RV samples
  have different means ($P_{\rm Stud}$), the significance level (in
  units of n-$\sigma$) that the two means are different following a
  Maximum-Likelihood approach (n-$\sigma_{\rm ML}$), and the
    number of radial bins used to search for rotation signals (N$_{\rm
      bin}$).}
\label{tab_vrot}
\end{table}

\newpage
\begin{table}[h!]
\begin{center}
%\begin{tabular}{lcccc}
\begin{tabular}{lrrrrrr}
\hline Cluster & $W_0$ & $c$ & $r_0$ & $r_c$ & $r_h$ & dist\\
\hline
NGC 288  & 5.80 & 1.21 & 76.56 & 70.0 & 190.00 &  8.83 \\
NGC 362  & 7.65 & 1.73 & 13.52 & 13.0 &  73.84 &  8.63 \\
NGC 1261 & 5.60 & 1.16 & 23.16 & 21.0 &  40.80 & 15.70 \\
NGC 1904 & 7.75 & 1.76 &  9.76 &  9.4 &  56.66 & 13.37 \\
NGC 3201 & 6.15 & 1.29 & 84.22 & 78.0 & 186.00 &  4.97 \\
NGC 5272 & 8.05 & 1.85 & 23.47 & 22.7 & 166.70 & 10.14 \\
NGC 5927 & 7.25 & 1.60 & 26.41 & 25.2 &  66.00 &  7.62 \\
NGC 6171 & 7.00 & 1.53 & 35.41 & 33.6 & 103.80 &  6.17 \\
NGC 6254 & 6.60 & 1.41 & 43.66 & 41.0 & 139.90 &  4.74 \\
NGC 6496 & 5.70 & 1.18 & 39.09 & 35.6 &  93.60 & 11.30 \\
NGC 6723 & 5.40 & 1.11 & 55.41 & 49.8 &  91.80 &  8.70 \\
\hline
\end{tabular}
\end{center}
\caption{Structural parameters of the King model that best fits the
  observed density/luminosity profile and cluster distances: King
  dimensionless potential and concentration parameter ($W_0$ and $c$,
  respectively), King, core and half-mass radii in arcseconds ($r_0$,
  $r_c$ and $r_h$, respectively), cluster distance in kpc (dist).  The
  structural parameters of NGC 6496 have been newly determined here,
  while the others are from \citet[][for NGC 288, NGC 1904, NGC 5272,
    and NGC 6254]{miocchi+13}, \citet[][for NGC 362]{dalessandro+13b},
  and \citet[][for NGC 1261, NGC 3201, NGC 5927, NGC 6171, and NGC
    6723]{harris96}. Distances are from \citet{ferraro+99} if
  available, otherwise they are from \citet{harris96}.}
\label{tab_struct}
\end{table}

\begin{table}[h!]
\begin{center}
\begin{tabular}{lccc}
\hline
Cluster  & $M$ & $M_{\rm McLvdM}$ & $M_{\rm B17}$ \\
\hline
NGC 288  &  0.78  &  0.74  &   0.88 \\
NGC 362  &  2.44  &  3.39  &   3.21 \\
NGC 1261 &  1.30  &   --   &    --  \\
NGC 1904 &  1.29  &   --   &   2.2  \\
NGC 3201 &  1.21  &  1.10  &   1.58 \\
NGC 5272 &  4.10  &  3.80  &   5.0  \\
NGC 5927 &  1.99  &   --   &   3.45 \\
NGC 6171 &  0.63  &  0.79  &   0.96 \\ 
NGC 6254 &  1.26  &  1.51  &   2.26 \\
NGC 6496 &  0.55  &  0.575 &    --  \\
NGC 6723 &  1.32  &  1.905 &   1.96 \\
\hline
\end{tabular}
\end{center}
\caption{Total mass of the program cluster in units of $10^5 M_\odot$
  as measured in the present paper ($M$), in \citet[][$M_{\rm
      McLvdM}$]{McLvdM05}, and in \citet[$M_{\rm B17}$]{baum17}.}
\label{tab_mass}
\end{table}

\end{document}